\documentclass[review,10pt,2022, 3p]{elsarticle}

\usepackage[utf8]{inputenc}
\usepackage[T1]{fontenc}
\usepackage{lscape}
\usepackage{caption}
\usepackage{subcaption}
\usepackage{booktabs}
\usepackage{comment} 
\usepackage{graphicx}
\usepackage[export]{adjustbox}
\usepackage{amsmath}
\usepackage{float}
\usepackage{graphics}

\usepackage{natbib}
\setcitestyle{authoryear,open={(},close={)}}

\usepackage{todonotes}      

\usepackage{hyperref}
\hypersetup{
    colorlinks      = {true},
    linkcolor       = {blue}, 
    linkbordercolor = {white},
    citecolor       = {blue},
    urlcolor        = {blue},
    pdftitle        = TourismReviewSLR2024
}

\usepackage{orcidlink}

\usepackage{doi}
    
\journal{arXiv}
\nocite{*}

\usepackage{tikz}
\usetikzlibrary{svg.path}
\definecolor{orcidlogocol}{HTML}{A6CE39}

\tikzset{
	orcidlogo/.pic={
		\fill[orcidlogocol] svg{M 100 100 C 100 72.4 77.6 50 50 50 C 22.3 50 0 72.4 0 100 C 0 127.6 22.3 150 50 150 C 77.6 150 100 127.6 100 100 Z};
		\fill[white] svg{M 33.666 77.27 L 27.651 77.27 L 27.651 119.103 L 33.666 119.103 L 33.666 100.198 L 33.666 77.27 Z};
		\fill[white] svg{M 42.493 119.103 L 58.742 119.103 C 74.21 119.103 81.007 108.049 81.007 98.167 C 81.007 87.425 72.609 77.23 58.821 77.23 L 42.493 77.23 L 42.493 119.103 Z M 48.509 82.66 L 58.078 82.66 C 71.71 82.66 74.835 93.011 74.835 98.167 C 74.835 106.565 69.484 113.673 57.766 113.673 L 48.509 113.673 L 48.509 82.66 Z};
		\fill[white] svg{M 34.603 127.813 C 34.603 125.665 32.846 123.868 30.658 123.868 C 28.471 123.868 26.713 125.665 26.713 127.813 C 26.713 130.001 28.471 131.758 30.658 131.758 C 32.846 131.758 34.603 129.961 34.603 127.813 Z};
	}
}

\def\orcid#1{%
	\href{https://orcid.org/#1}{%
		\begin{tikzpicture}[baseline]
			\pic[scale=0.1,yshift=-7em] {orcidlogo};
		\end{tikzpicture}
	}%
}

\NewDocumentCommand{\plot}{m O{} O{12} O{H} O{trim=0cm 0cm 0cm 0cm} }{
\begin{figure}[#4]
\centering
\includegraphics[width=#3cm,clip,#5]{#1}
\caption{#2}
\label{#1}
\end{figure}
}

\newcommand{\orcidd}[1]{\href{https://orcid.org/#1}{\textcolor[HTML]{A6CE39}{\aiOrcid}}}

\newcommand{\blindreview}[1]{\ifblind\textbf{[Blinded for review]}\else#1\fi}

\newif\ifblind

\begin{document}

\begin{frontmatter}

\title{Digital twins in tourism: a systematic literature review}

\ifblind
\else
    \author[1]{Duarte Sampaio de Almeida\orcidlink{0000-0001-5459-4113}}
    \ead{dsbaa@iscte-iul.pt}
    
    \author[1]{Fernando Brito e Abreu\orcidlink{0000-0002-9086-4122}}
    \ead{fba@iscte-iul.pt}
    
    \author[2,3]{Inês Boavida-Portugal\orcidlink{0000-0001-9932-9241}}
    \ead{iboavida-portugal@campus.ul.pt}

    \affiliation[1]{
        organization={Information Sciences, Technologies and Architecture Research Centre, Iscte - Instituto Universitário de Lisboa},
        country={Portugal}
    }
    
    \affiliation[2]{
      organization={Centre of Geographical Studies, Institute of Geography and Spatial Planning, University of Lisbon},
      country={Portugal}
    }
    
    \affiliation[3]{
      organization={Associated Laboratory Terra},
      country={Portugal}
    }
\fi

\begin{abstract}

\noindent{\textbf{Purpose:}} This systematic literature review (SLR) characterizes the current state of the art on digital twinning (DT) technology in tourism-related applications. We aim to evaluate the types of DTs described in the literature, identifying their purposes, the areas of tourism where they have been proposed, their main components, and possible future directions based on current work.

\noindent{\textbf{Design/methodology/approach:}}
We conducted this SLR with bibliometric analysis based on an existing, validated methodology.
Thirty-four peer-reviewed studies from three major scientific databases were selected for review. They were categorized using a taxonomy that included tourism type, purpose, spatial scale, data sources, data linkage, visualization, and application.

\noindent{\textbf{Findings:}} The topic is at an early, evolving stage, as the oldest study found dates back to 2021. Most reviewed studies deal with cultural tourism, focusing on digitising cultural heritage. Destination management is the primary purpose of these DTs, with mainly site-level spatial scales. In many studies, the physical-digital data linkage is unilateral, lacking twin synchronization. In most DTs considered bilateral, the linkage is indirect. There are more applied than theoretical studies, suggesting progress in applying DTs in the field. Finally, there is an extensive research gap regarding DT technology in tourism, which is worth filling.

\noindent{\textbf{Originality/Value:}} This paper presents a novel SLR with a bibliometric analysis of DTs' applied and theoretical application in tourism. Each reviewed publication is assessed and characterized, identifying the current state of the topic, possible research gaps, and future directions.

\end{abstract}


\end{frontmatter}

\section{Introduction}
\label{Introdution}

The term ``digital twin'' (DT) was coined on May 10, 2011, in a \href{https://www.nasa.gov/}{NASA} workshop focused on Modeling, Simulation, Information Technology, and Processing. During the discussion on the simulation of engineering systems, \href{https://www.aiaa.org/detail/person/greg-zacharias}{Greg Zacharias} highlighted the concept of complete end-to-end systems modeling as a transformative approach to systems engineering. He referred to this comprehensive modeling capability, incorporating human operators with appropriate cognitive and motor fidelity, as a ``Digital Twin'' \citep{NASA2012}. This marked an early articulation of the concept, emphasizing its potential to revolutionize how systems are designed, tested, and operated by integrating detailed, multi-resolution models of machines and humans.

Several definitions of a DT have been proposed in the literature, such as the nice and clean short one provided by the \cite{AIAA2020}: \textit{``A Digital Twin is a virtual representation of a connected physical asset.''} Detailing a bit more, a DT is a dynamic digital representation of an entity or system that mirrors its physical counterpart's structure, context, and behavior. It is continuously updated with real-time data from the physical system (physical-to-digital connectivity) and, in turn, enables monitoring, prediction, simulation, and decision-making that can influence the physical counterpart (digital-to-physical connectivity). By maintaining these bidirectional interactions throughout the asset's lifecycle, a digital twin enhances understanding, optimization, and value realization.

\begin{figure}[H]
    \centering
    \includegraphics[width=1\textwidth]{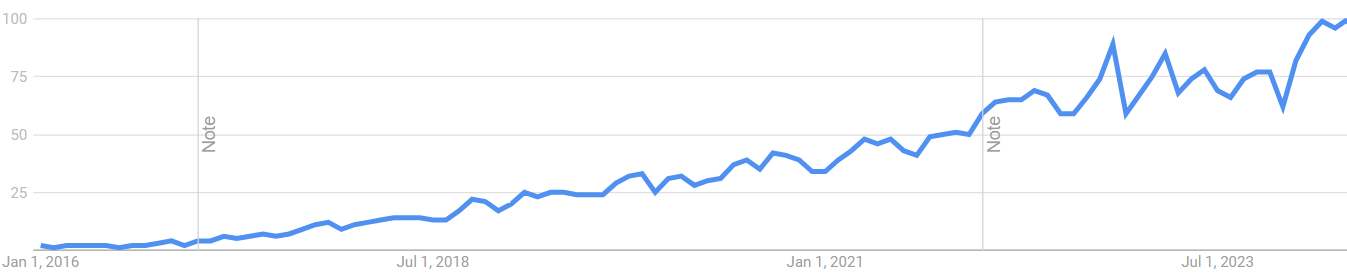}
    \caption{Worldwide interest over time on digital twins from 2016 through May 2024 (Source: Google Trends)}
    \label{fig:dt_relevance}
\end{figure}

Even though the concept of DT has existed for more than a decade, its relevance in several domains (e.g., industry, health, engineering) has gained momentum in recent years. A clear sign of this was the creation of the \href{https://www.digitaltwinconsortium.org/}{Digital Twin Consortium}, a collaborative partnership with industry, academia, and government expertise, led by the \href{https://www.omg.org/}{Object Management Group}, to drive the awareness, adoption, interoperability, and development of digital twin technology. The growth of DT relevance is also evident in \href{https://trends.google.com/}{Google Trends}, as shown in Figure \ref{fig:dt_relevance}.

The tourism industry is also a domain where DT technology is starting to grow, following the success of applying smart tourism tools in diverse areas such as tourism overcrowding management \citep{Abreu2024} and many more, either as part of the tourist offer, for marketing purposes or tourism management and operations \citep{Galvao2024}.

Although several primary studies on the application of DTs in tourism have been published, to the best of our knowledge, there is no state-of-the-art review on the topic at the time of writing. Recent literature reviews on DTs barely mention its use in tourism-related applications, at least directly. The closest secondary study we have come across is \cite{Dang2023}, which examines the application of DTs to world heritage sites in China. However, the latter focuses on digitizing heritage for conservation purposes rather than the tourism phenomenon that often results from the fruition of that heritage. 

Our main objective is to conduct a systematic literature review (SLR) to characterize the current state of DT technology implemented in tourism-related applications. We aim to evaluate the types of DT described in the literature, identifying their purposes, namely in which areas of tourism they have been proposed and their main components, and outline future directions for this topic.

This paper is organized as follows: in section \ref{Background}, we introduce key aspects of DT studies; in section \ref{RelatedWork}, we review some of the most relevant related works; in section \ref{Methodology}, we present the methodology of this SLR; in section \ref{Results&Discussion} we present the results of our study; finally, in section \ref{Conclusions&FutureWork} we draw some conclusions.

\section{Background}
\label{Background}

In the aerospace context, where the term was coined in 2011, the definition of DT evolved to \textit{``an integrated multiphysics, multiscale, probabilistic simulation of an as-built vehicle or system that uses the best available physical models, sensor updates, fleet history, etc., to mirror the life of its corresponding flying twin''} \citep{Glaessgen2012}.

Since then, many authors have provided non-consensual definitions for DTs \citep{Barricelli2019, Haag2019}. Although often overlapping, because most consider that ``a Digital Twin consists of three parts: physical product, virtual product, and connected data that tie the physical and virtual product'' \citep{Tao2019}, DT definitions are highly dependent on individual use cases \citep{Wagner2019}. More recently, \cite{vanderValk2022} proposed a ranking taxonomy for DTs based on archetypes, each including a set of mandatory and non-mandatory characteristics.

In recent years, DT technology has spread from aerospace and manufacturing to many other domains such as healthcare \citep{Willcox2024}, smart cities \citep{Wan2023}, or the Earth as a whole \citep{Hazeleger2024}. Regarding the latter, it is worth mentioning the European Commission's flagship initiative -- \href{https://destination-earth.eu/}{Destination Earth (abbreviated \emph{DestinE})} -- to develop a highly accurate digital twin of the Earth to model, monitor, and simulate natural phenomena, hazards, and human activities. These groundbreaking features assist users in designing accurate and actionable adaptation strategies and mitigation measures.

Tourism is no exception. In recent years, this sector has embraced digital transformation \citep{Bekele2024}. An increasingly significant phenomenon affecting tourism nowadays is the excessive affluence of tourists to hotspots around the globe, degrading ecosystems, the tourist experience, and the quality of life of residents of those hotspots. The World Tourism Organization \citep{UNWTO2018} defines this as \textit{overtourism}. The Spanish city of Barcelona is a well-known example of the effects of unsustainable tourism, where a major tourism-phobia movement is taking place \citep{PirilloRamos2021}. 
With the emergence of overtourism, it is more relevant than ever to monitor and predict tourist behavior to apply mitigation measures for this phenomenon effectively. The monitoring and prediction capabilities of DTs have the potential to provide valuable insights into the complex tourism behavior system, allowing decision makers to apply better-informed action plans.

At the other end of the spectrum, during the COVID-19 pandemic, the tourism system took a massive hit, as travel and most tourism activities were suspended to prevent the spread of the pandemic \citep{Casal2023}. Many traditional outdoor processes were digitalized as people were forced to stay home. One of the boosted digital evolutions was virtual tourism \citep{Verma2022}, consisting of the virtual visitation and exploration of digital representations of tourism destinations, usually with the support of Extended Reality (XR) technologies. This allowed people to explore scenic tourist spots from their homes during such restrictive times. DT technologies help create realistic, immersive virtual tourism experiences, as the high fidelity of digital counterparts and the data linkage between physical and digital allows for sophisticated virtual interactions. This could also be known as Metaverse, loosely referring to digital worlds explorable through XR technologies. This set of technologies has great potential to revolutionize the tourism industry, facilitating tourist engagement, remote exploration, and intention of actual, physical visitation \citep{Buhalis2024}. As such, there are evident benefits in adopting DT technology in tourism.

\section{Related work}
\label{RelatedWork}

Some Systematic Literature Reviews (SLR) and state-of-the-art surveys have recently been performed to review the body of knowledge in DT technology. We will address review studies in which the authors search for multiple domain applications of DT (instead of, e.g., reviews of DT in specific fields, such as manufacturing or healthcare) published after 2019.

\cite{Barricelli2019} present a survey on the state-of-the-art of DT technology, where they investigate its definitions, the main characteristics a DT should possess, and the domains in which DT applications are being developed. The survey considered seventy-five documents, including peer-reviewed and grey literature.
The domains of the addressed papers were manufacturing, aviation, hospital management, and precise medicine, but not tourism. The authors recognize the viability of using DT in other domains.

\cite{Semeraro2021} presented an SLR based on a broad systematic literature review on DT studies, tools, and technicalities, focusing on the definition of DT, their characteristics, and design implications. The authors aimed to trace the ongoing DT research and technical challenges in conceiving/building them according to different application domains and technologies. After selection, the authors identified and analyzed 150 peer-reviewed papers. Using text mining and clustering techniques, the authors extracted features related to the definitions and topics of DT. The existing DT applications were divided into the following categories: Healthcare, Maritime and Shipping, Manufacturing, City Management, and Aerospace.
This study presented the application contexts, life cycle phases, functionalities, architectures, and components of DTs.

\cite{BotinSanabria2022} conducted an SLR that presents a comprehensive view of the DT technology, its implementation challenges, and limits in the most relevant domains and applications. Eighty-four publications were obtained for presentation, with a smaller set of 18 more recent (2019–2022) studies for a deeper comparative analysis to present insights into the trending enabling technologies used in DT-based systems. The application domains were the following: smart cities and urban spaces, freight logistics, medicine, engineering, and automotive.

\cite{Jones2020} presented an SLR and a thematic analysis of the characterization of the DT, identification of gaps in knowledge, and future work directions. Ninety-two publications were considered for analysis.
A total of thirteen characteristics emerged: Physical Entity/Twin; Virtual Entity/Twin; Physical Environment; Virtual Environment; State; Realisation; Metrology; Twinning; Twinning Rate; Physical-to-Virtual Connection/Twinning; Virtual-to-Physical Connection/Twinning; Physical Processes; and Virtual Processes. 
Seven knowledge gaps and topics for future research were identified: Perceived Benefits; Digital Twin across the Product Life cycle; Use cases; Technical Implementations; Levels of Fidelity; Data Ownership; and Integration between Virtual Entities.

\cite{Hu2021} presents a state-of-the-art review of the enabling technologies, applications, and challenges of developing DTs.
This work reviews six categories of enabling technologies, definitions, models, and an overview of the history of DT. Additionally, it offers a thorough analysis of DT applications from two angles: Applications in four stages of the product lifecycle and (2) applications in four engineering fields: wind, Internet of Things (IoT) applications, tunnelling and subterranean engineering, and aerospace engineering. This study extracts the DT frameworks, characteristic components, essential technologies, and particular applications for each DT category.
Through the review, the authors concluded that most DT models only involve unilateral data linkage (from physical to virtual twins), and the environmental coupling is often overlooked, resulting in inaccurate digital representations. \cite{Heluany2023} conducted an SLR on DT reviews (tertiary study). Its goal is to understand the use cases, modelling and simulation tools/techniques, and security implementation. The authors conclude that there is a misconception and misuse regarding the DT concept, and security is often neglected but mentioned as a challenge. 

The identified literature reviews of DT technology do not directly mention tourism. As such, literature reviews on the application of DT in tourism-related scenarios seem to be a research gap to fill.


\section{Methodology}
\label{Methodology}
As aforementioned, this work is an SLR. As opposed to an unstructured (ad-hoc) review, this methodology aims to reduce bias and systematize the review process by following a rigorous chain of steps to search, extract, analyze, and present results, allowing for its replication and mitigating validity threats \citep{Prisma2020}. The methodology used in this work is based on the one presented by \cite{CarreraRivera2022}. This section describes these methodological steps.

\subsection{Research questions}

The following questions drive this research:
\begin{itemize}
    \item \textbf{RQ1}: How developed is the state-of-the-art of digital twinning in tourism?
    \item \textbf{RQ2}: How are digital twins used for tourism-related applications? 
    \item \textbf{RQ3}: What are the research gaps and future directions of digital twins for tourism?
\end{itemize}

\subsection{Scientific databases}
The topic addressed in this review spans different scientific domains (e.g., social sciences and information technology). As such, we opted to search for related work on interdisciplinary, broad-search scientific databases with wide coverage of indexed peer-reviewed studies, with the option of searching the selected strings in the topic of the studies (i.e., title, abstract, and keywords).
The chosen ones were the following:

\begin{itemize}
    \item \href{https://www.scopus.com/}{Scopus};
    \item \href{https://www.webofscience.com/}{Web of Science};
    \item \href{https://app.dimensions.ai/}{Dimensions}.
\end{itemize}

\subsection{Publications time frame}
As tourism-applied digital twinning is a very recent topic, we opted not to limit the search to a specific time frame, as the search results naturally only include recently published studies.

\subsection{Search Strings}

In the databases that support wildcards (i.e., Scopus, Web of Science), the search string used was the following:
\begin{itemize}
    \item \textit{"digital twin" and "touris*"}
\end{itemize}

\noindent As Dimensions does not allow the usage of wildcards in the free version, the search string was the following:
\begin{itemize}
    \item \textit{"digital twin" and (tourism OR tourist)}
\end{itemize}

The search strings were searched in the title, abstract, and keywords (also known as \textit{topic}), focusing the search on the studies where DT and tourism are considered key topics. 

\subsection{Inclusion and exclusion criteria}
To include or exclude found articles in this review, the defined criteria are presented in Table \ref{tab:inclusion_exclusion}.

\begin{table}[H]
\centering
\small
\caption{Inclusion and Exclusion Criteria (Source: Authors own work)}
\label{tab:inclusion_exclusion}
\begin{tabular}{ll} 
\toprule
Criterion & Description \\
\midrule
\multicolumn{2}{l}{\textbf{Inclusion criteria}} \\
IC1          & Studies in open access format or available through our institutions \\        
IC2          & Peer-reviewed studies from any venue \\    
IC3          & Works that present implemented or theoretical applications of DT in tourism systems \\   
\multicolumn{2}{l}{\textbf{Exclusion criteria}} \\
EC1          & Weak connection to DT or tourism \\        
EC2          & Studies not written in English;  \\    
EC3          & Secondary or tertiary studies \\      
\bottomrule
\end{tabular}
\end{table}

\subsection{Selection process}

The executed selection process steps are the following:
\begin{enumerate}
    \item 
    Apply the search strings on their respective databases, obtaining possibly relevant studies automatically;
    \item Remove duplicate references; 
    \item Apply the inclusion criteria (orderly by title, abstract, keywords, and content) on the studies obtained;
    \item Apply the exclusion criteria;
\end{enumerate}

\begin{figure}[H]
    \centering
    \includegraphics[width=1.0\textwidth]{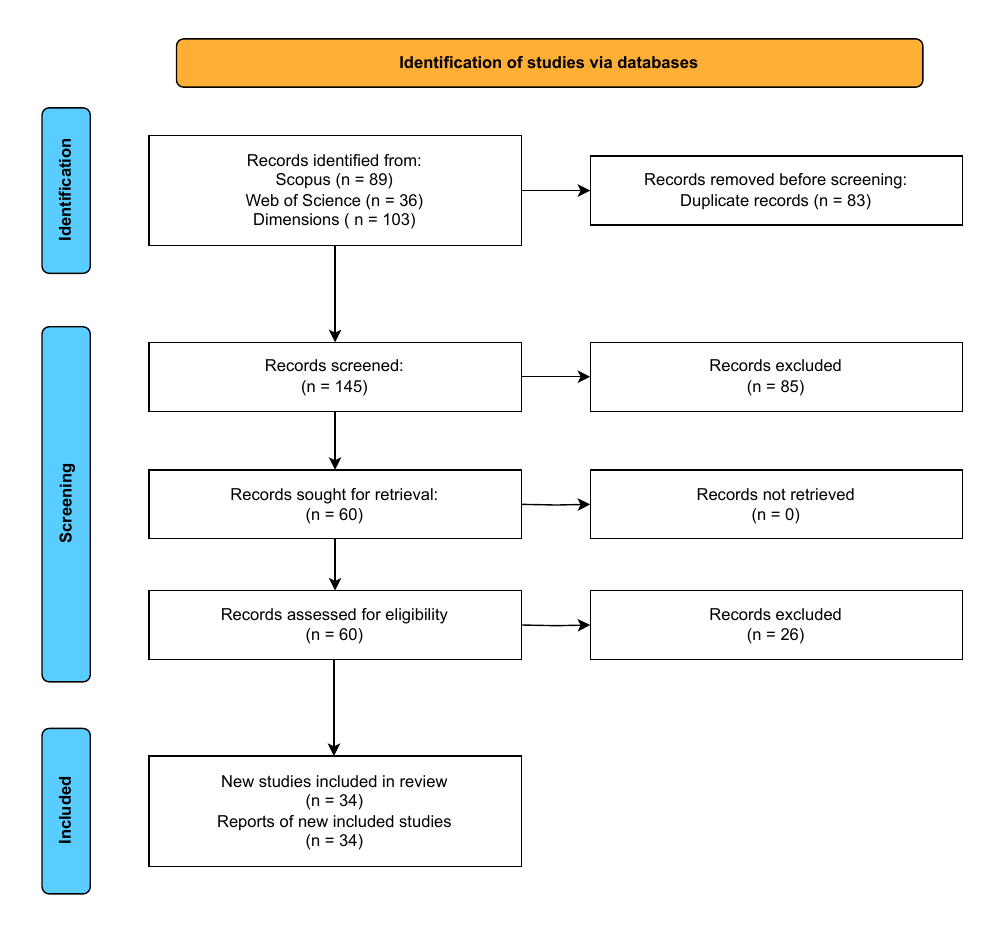}
    \caption{PRISMA Flow Diagram of the study selection process (Source: Authors own work)}
    \label{fig:prisma}
\end{figure}

Figure \ref{fig:prisma} illustrates the previously described process using a PRISMA flow diagram \citep{Prisma2020}.

\noindent

The last article search on the selected databases was conducted on December 1, 2024. At this stage, 228 references were collected, with 83 duplicated and 111 rejected references, leaving a selection of 34 publications to be considered for review.

\subsection{Classification taxonomy}
\subsubsection{Taxonomy criteria}
A set of criteria was defined to categorize the different types of tourism-applied DT studies. The possible values for each criterion were obtained from study analysis and represent all types found in the reviewed studies. The criteria were selected based on the main components of digital twins (data sources, linkage, application, visualization), purposes/tourism application domain (purpose, tourism type), and spatial scope (spatial scale), as tourism services and subsystems can scope different spatial scales. Studies are classified with the value "N/A" in any criteria if the information is not extracted from the publication or the requirements are not applicable.
The following subsections present the chosen set of criteria.

\subsubsection{Tourism type}
The studies on tourism DTs can be categorized by the type of tourism activity and the environment where they are applied. This criterion lists the tourism categories where the DT technology is applied, being defined on a nominal scale with the following overlapping categories:
\begin{itemize}
    \item \textbf{Generic (G)}: The type of tourism is not explicitly described or fixed into a specific type, as the application may be theoretical. The DT could be applied to multiple types of tourism.

    \item \textbf{Cultural (C)}: Visiting cultural heritage sites, such as monuments, heritage buildings, and art.
    
    \item \textbf{Natural (N)}: Contact with nature and exploring geologic, ecologic, botanic, and/or zoological interest areas.
    
    \item \textbf{Rural (R)}: Visiting rural areas.  
    
    \item \textbf{Urban (U)}: Visiting urban environments (e.g., historical neighbourhoods, city squares, urban monuments).

    \item \textbf{Science (S)}: Visiting scientific attractions, such as science museums, labs, and scientific activities. 

    \item \textbf{Culinary (Cu)}: Partaking in culinary and food experiences and trying local cuisines from visited destinations.
    
\end{itemize}

\subsubsection{Purpose}
DT technology has a wide range of useful applications in tourism scenarios. This criterion lists the main purposes the DTs discussed in the reviewed works have, being defined on a nominal scale with the following overlapping categories:
\begin{itemize}
    \item \textbf{Digital preservation (DP)}: These DTs are used to preserve objects of touristic or historical interest digitally, such as 3D models and historical documents;

    \item \textbf{Virtual tourism (VT)}: The DTs provide the exploration of destinations in virtual worlds in a computerized medium;

    \item \textbf{Experience enhancement (EE)}: This DT allows for the virtual enhancement of the physical tourism experience. AR technology and tourism activity recommendations are fine examples.

    \item \textbf{Destination management (DM)}: The DTs classified with this category aid stakeholders in managing the tourism destination.
    
\end{itemize}

\subsubsection{Spatial scale}
DTs can have different scales regarding spatial coverage and detail grain, from single sites to villages/cities, regions, countries, etc. The nominal scale representing the overlapping criterion is the following:
\begin{itemize}
    \item \textbf{Site (S)}: The DTs represent tourism sites, such as single monuments, parks, buildings, etc.

    \item \textbf{Local (L)}: The DTs represent wider-scoped geographies than the Site level. Systems that describe towns, villages, or cities have this scale.

    \item \textbf{Regional (R)}: The spatial scale above Local. These DTs describe regional-level tourism systems, scoping sets of locals, regions, provinces, districts, etc.
\end{itemize}

\subsubsection{Data sources}
This overlapping criterion describes the data sources used as input for the DT.
The nominal scale representing the criterion is the following:
\begin{itemize}
    \item \textbf{3D scanning (3DS)}: The data is obtained from specialized technologies that can capture and create a 3D digital representation of a physical entity;
    \item \textbf{Geographic Information Systems (GIS)}: This data source consists of geo-referenced data, usually from mapping databases.
    \item \textbf{Sensing (S)}: The DT is fed by sensors of any kind. These sensors can capture dynamic changes in the physical entity, such as human activity and environmental data.
    \item \textbf{User-generated Data (UGC)}: The data is sourced through user-generated content, such as tourism site reviews, social media data, etc.
    \item \textbf{Web mining (WM)}: This source captures data from the web through scraping and other techniques.
    \item \textbf{Historical data (HD)}: The data is obtained from historical artifacts (e.g., documents, works of art) containing information about the physical entity. 
    \item \textbf{Environmental data (ED)}: This source served data about the environment where the entity is located. It may come from sensors or other sources, such as weather entities.
    \item \textbf{AI-generated data (AIGD)}: The data is created with generative AI tools (e.g., ChatGPT, Dall-E, Claude)
    \item \textbf{Medical data (MD)}: This source adds health components to DTs, such as disease infection data and mortality data in vehicular accidents.
    \item \textbf{Socio-economic data (SED)}: This source provides data about the social and economic factors of the population living near or in the tourism destination. Census data is an example.
    \item \textbf{Surveying (Su)}: The data is obtained from surveys and questionnaires conducted with people.
\end{itemize}

\subsubsection{Data linkage}
This mutually exclusive criterion describes the data linkage between digital and physical twins. 
The ordinal scale representing the criterion is the following:
\begin{enumerate}
    \item \textbf{Unilateral (U)}: The connection between twins is half-duplex, unidirectional (from physical to digital). The data from the digital counterpart does not have a clear direct or indirect impact on the physical entity.
    \item \textbf{Bilateral (B)}: Both twins connect with each other. The digital and real entities synchronize with each other. We consider indirect connections from the digital to the physical (e.g., the DT results can influence and possibly induce changes to the physical counterpart, which in turn change the digital).
\end{enumerate}

\subsubsection{Visualization}
This overlapping criterion describes the visualization and Human-Machine Interface (HMI) technologies used for the DT applications in the studies, being represented in a nominal scale by the following values:
\begin{itemize}
    \item \textbf{Augmented Reality (AR)}: This technology virtually enhances physical reality with computer graphics. 
    \item \textbf{Virtual Reality (VR)}: The technology allows immersion in virtual 3D worlds, which can be faithful representations of the physical world, altered versions, or wholly fabricated worlds.
    \item \textbf{Web (W)}: The DT can be visualized and/or interacted with through a standard web application without special visualization technologies (e.g., VR, AR).
    \item \textbf{Mobile app (MA)}: The DT can be visualized through a mobile application without special visualization technologies.
    \item \textbf{Projection Mapping (PM)}: This technology involves projecting video or image onto an often irregular surface. This can add interest to the surface or dimensions to a flat video or image.
    \item \textbf{Desktop (D)}: The DT can be visualized through a desktop application without special visualization technologies.
\end{itemize}

\subsubsection{Application}
The mutually exclusive criterion lists if the DT study is applied in real-life scenarios, being deployed and ready-to-use, or prototypes assessed through case studies, or if it is a theoretical framework not yet implemented or tested in a case study. An ordinal scale represents it:
\begin{enumerate}
    \item \textbf{Theoretical (T)}: The study is theoretical and was not applied at the time of its publication.
    \item \textbf{Applied (A)}: The study is applied in real-life scenarios. We consider "Applied" if the DT has an implemented prototype.
\end{enumerate}

\section{Results and discussion}
\label{Results&Discussion}

\subsection{Bibliometric analysis}

\begin{figure}[H]
    \centering
    \includegraphics[width=0.7\textwidth]{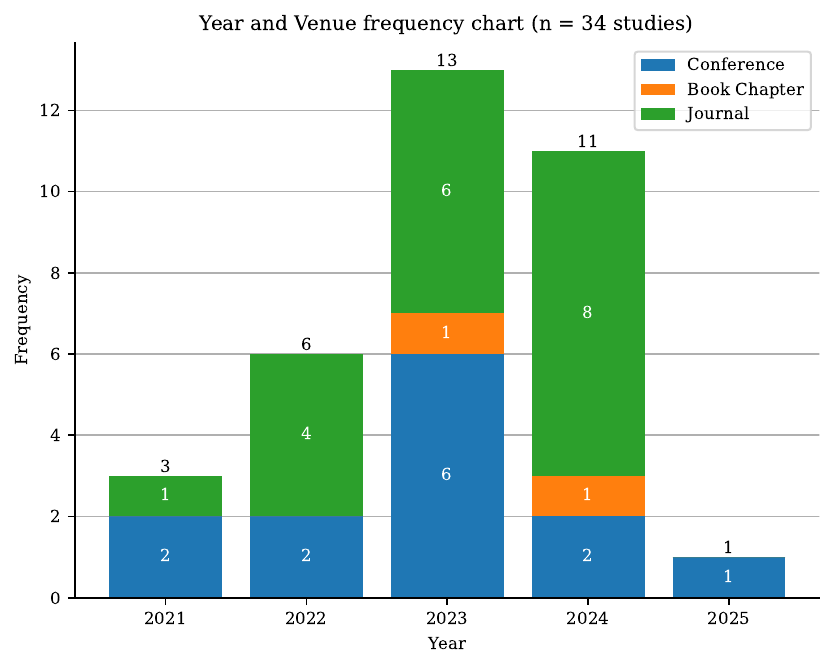}
    \caption{Number of publications by year, subdivided by venue (Source: Authors own work)}
    \label{fig:publications_year}
\end{figure}

In Figure \ref{fig:publications_year}, we can observe how the number of publications presenting or proposing DTs for tourism has increased over time. Although the earliest studies were found in 2021, and the number of studies is low, indicating the recency of this topic, the number of studies has increased each year (until 2023, as 2024 was not a complete year when the last search was conducted). There is one publication with the publication year 2025, but it should be considered as from 2024.
We can also observe the number of peer-reviewed publications by venue each year. Most studies are published in journals, but they are also published in conferences and book chapters. The publication of these studies is evenly scattered by different publication sites. Almost every study was published on a different site. As such, there is not a dominating one yet.

\begin{figure}[H]
    \centering
    \includegraphics[width=0.5\textwidth]{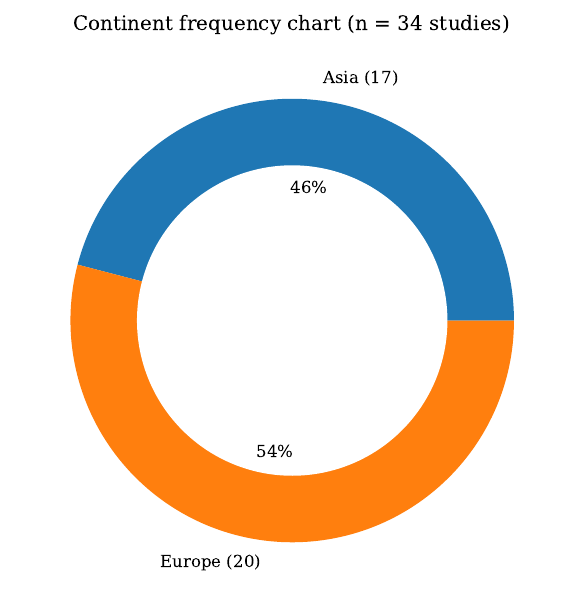}
    \caption{Continent distribution histogram (Source: Authors own work)}
    \label{fig:continents}
\end{figure}

\begin{figure}[H]
    \centering
    \includegraphics[width=0.85\textwidth]{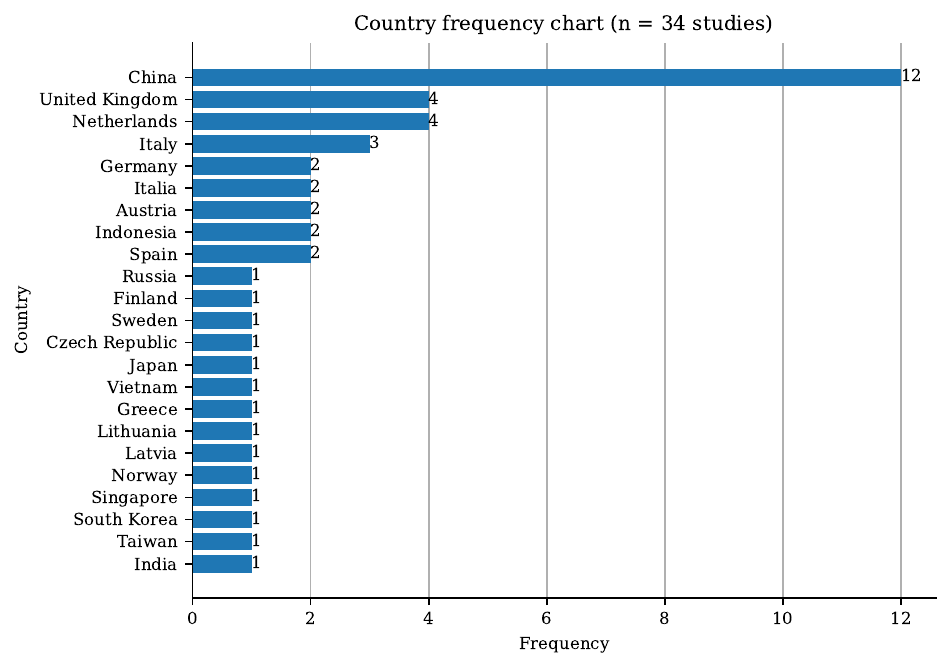}
    \caption{Country frequency (Source: Authors own work)}
    \label{fig:countries}
\end{figure}

Figures \ref{fig:continents} and \ref{fig:countries} show, respectively, the distributions of the continents and countries involved in the studies (obtained from author affiliations). Tourism DT studies are being conducted and presented by authors from Europe and Asia with almost equal contributions (Europe in 54\% of the studies, Asia in 46\%). Other continents did not contribute to the reviewed studies in authorship.
A total of 23 countries contributed to the selected studies. China has the most significant contribution to the reviewed body of knowledge, being involved in $N = 12$ studies. The United Kingdom and the Netherlands are the second most contributing countries, involved in $N = 4$ studies each. Other countries have also made similar contributions to these studies.

\subsection{Study analysis}

\subsubsection{Tourism type}
\begin{figure}[H]
    \centering
    \includegraphics[width=0.75\textwidth]{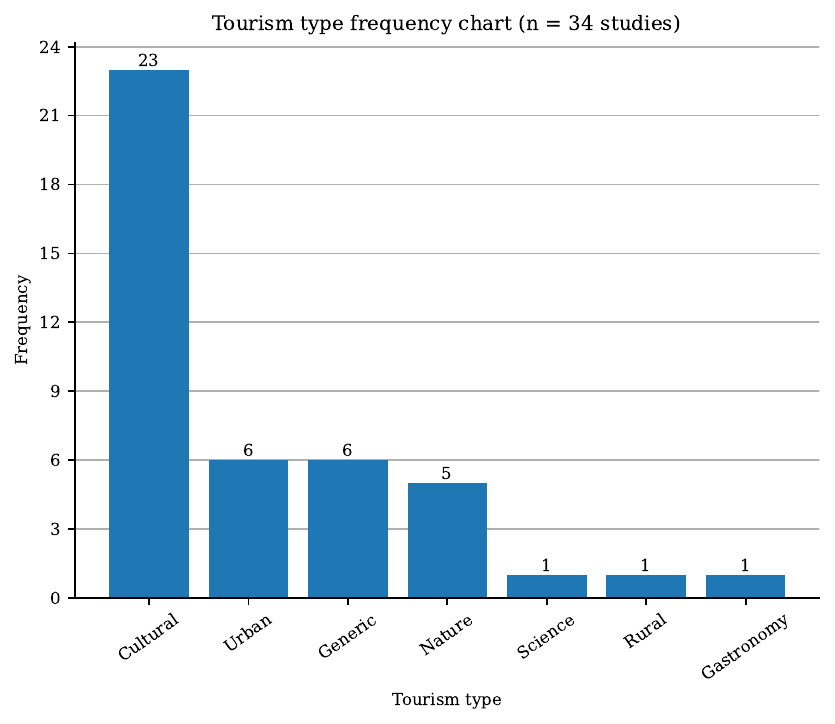}
    \caption{Frequency of tourism type categorizations in studies (Source: Authors own work)}
    \label{fig:tourism_types_freq}
\end{figure}

Figure \ref{fig:tourism_types_freq} depicts the frequency of studies categorized in each tourism type dimension. Most studies are applied to cultural tourism ($N = 23$). These studies can be subcategorized into generic cultural heritage \citep{Chen2023}, cultural ecosystems, joining culture and natural tourism \citep{Rolph2024}, monuments \citep{Zhang2022}, art \citep{Zhao2022, Fistola2023}, heritage buildings \citep{Doria2023, Nguyen2022, LopezGonzalez2023, Kumar2022, Guzzetti2022, Banfi2022, Tang2024, Galiano2024, Dayoub2024}, archaeology \citep{Chen2024} and entire villages, cities or regions \citep{Gallist2023, Yang2021, Sang2022, Zhou2023, Dayoub2024, Kourtit2025, Planu2024, Huang2024, Feklistov2024}, having some variety in this application type. 
These studies typically focus on digitalizing intact or damaged heritage into 3D interactive models with different levels of detail for virtual preservation, restoration, reimagination, structural analysis, and exploration by cyber-tourists. This allows them to experience these landmarks anyplace in the world, with new twists and perspectives.

$N = 6$ studies can be applied to generic tourism, being possible to be implemented in various types \citep{Zhang2023, Rahmadian2023Governing, Rahmadian2023, Litavniece2023, Deng2024, Prabawati2024}. Two of these publications address the same study, but with different details, being published in different venues \citep{Rahmadian2023, Rahmadian2023Governing}. The studies classified as "generic" focus on providing theoretical frameworks or guidelines to apply DT technology to a wide scope of tourism objects or specific steps in developing DTs, such as data governance.

$N = 6$ studies present DTs of urban tourism scenarios, where digital versions of parts of a city with tourism interest are presented. \cite{Gallist2023} focus on creating a digital version of a city in the Alps, explorable through VR in a Metaverse application. \cite{Yang2021} presented a theoretic DT of the city of Guilin for cultural tourism. \cite{LopezGonzalez2023} presented a DT-based platform for cultural heritage and urban sites of interest. \cite{Li2024} presented an urban DT that describes socio-economic situations relevant to tourism, such as security, economy and epidemic prevention and control. \cite{Fistola2023} present the Vox Hortus, a DT system describing urban artworks of interest and adding interactive layers (e.g., sound) and using XR to enhance their real-world visitations.

$N = 5$ studies are also applied to nature tourism. In this type of application, natural life \citep{Boletsis2022}, environment and ecologic features \citep{Boletsis2022, Logothetis2023, Rolph2024}, natural structures and geologic features \citep{FischerStabel2021} are digitally represented. \citep{Planu2024} present a DT that use various types of tourism, including natural.
These DTs also allow users to explore the natural landscapes virtually or enhance the visitation of the physical environment with additional interactive or static information.

The remaining types extracted were rural tourism \citep{Sang2022}, where the main activity is visiting rural places of interest, science tourism \citep{Seungyoub2021}, where tourism activities have scientific interest, and culinary tourism \citep{Planu2024}, focused on food-related activities.

\subsubsection{Purpose}
\begin{figure}[H]
    \centering
    \includegraphics[width=0.75\textwidth]{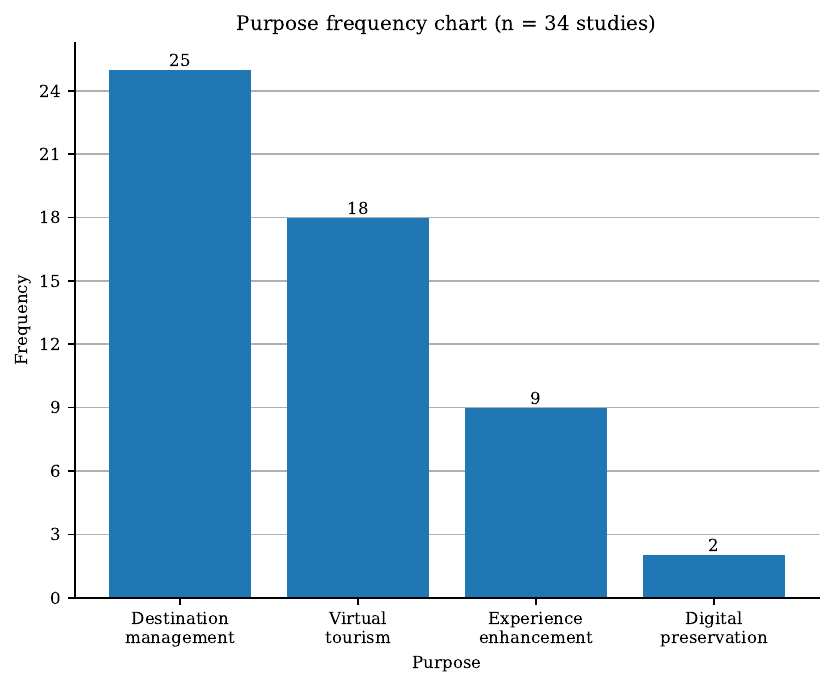}
    \caption{Frequency of purpose categorizations in studies (Source: Authors own work)}
    \label{fig:purpose_freq}
\end{figure}

We defined the purposes of the reviewed DTs in 4 high-level categories specific to tourism.
Figure \ref{fig:purpose_freq} depicts the frequency of classified studies for each purpose. Destination management is a purpose of most reviewed DTs $(N = 25)$. These DTs offer features to practitioners to help them manage tourism destinations, such as monitoring the state of tourism activity and consumer satisfaction, simulating action response and destination preservation scenarios, and improvement. 
Virtual tourism is another widely used DT purpose ($N = 18)$. As aforementioned, this system allows the virtual exploration of faithful or reimagined digitalized destinations in virtual worlds, with its inherent advantages. 
The third purpose is experience enhancement $(N = 9)$. These systems allow tourists to enhance their activities through activity prescription (e.g., route recommendation) or adding digital information or interaction layers to physical activities (presented through AR). The final purpose is digital preservation $(N = 2)$. These DTs store physical objects of touristic interest (e.g., artwork) in repositories but do not provide interfaces for other purposes.
Many reviewed studies have multiple purposes, harnessing the underlying DT systems to their fullest. 

\subsubsection{Spatial scale}

\begin{figure}[H]
    \centering
    \includegraphics[width=0.75\textwidth]{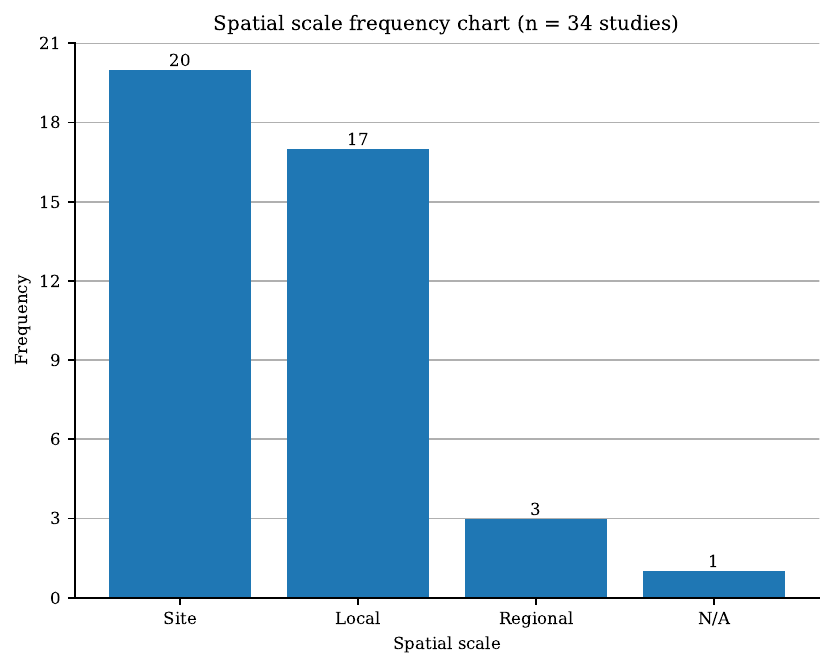}
    \caption{Frequency of spatial scale categorizations in studies (Source: Authors own work)}
    \label{fig:spatial_scale_freq}
\end{figure}

Figure \ref{fig:spatial_scale_freq} depicts each spatial scale's frequency of classified studies. The spatial scale of a DT usually reflects the amount of data it needs. The larger the scale, the more data is expected to power its representative DT. Moreover, the more data is ingested, the more computational power will be needed for a usable experience. This can be mitigated by pre-processing data at the expense of reducing detail. As such, even though larger-scale systems have more data, the grain should be higher, focusing on larger-scale processes and entities.

Almost all reviewed studies have site and local spatial scales. These represent tourism sites, sets of sites, or local zones of touristic interest (e.g., towns, municipalities, and cities). These scales are suitable for every main tourism DT purpose, as the smaller dimension of the described system allows data to be fine-grained and, therefore, used to create detailed descriptions of physical spaces and tourism dynamics.
These scales are better suited to local-scale practitioners and tourism site managers.
A few studies present DTs with regional spatial scales, which are more suited for governmental stakeholders, global tourism operators/entities, and researchers studying large-scale tourism systems. These scales allow for a less-detailed, higher-grained representation of these systems. The major caveat in developing systems with such a scale is the increased complexity of the system, as it may scope numerous complex subsystems and the relationships between them.

\subsubsection{Data sources}
\begin{figure}[H]
    \centering
    \includegraphics[width=0.75\textwidth]{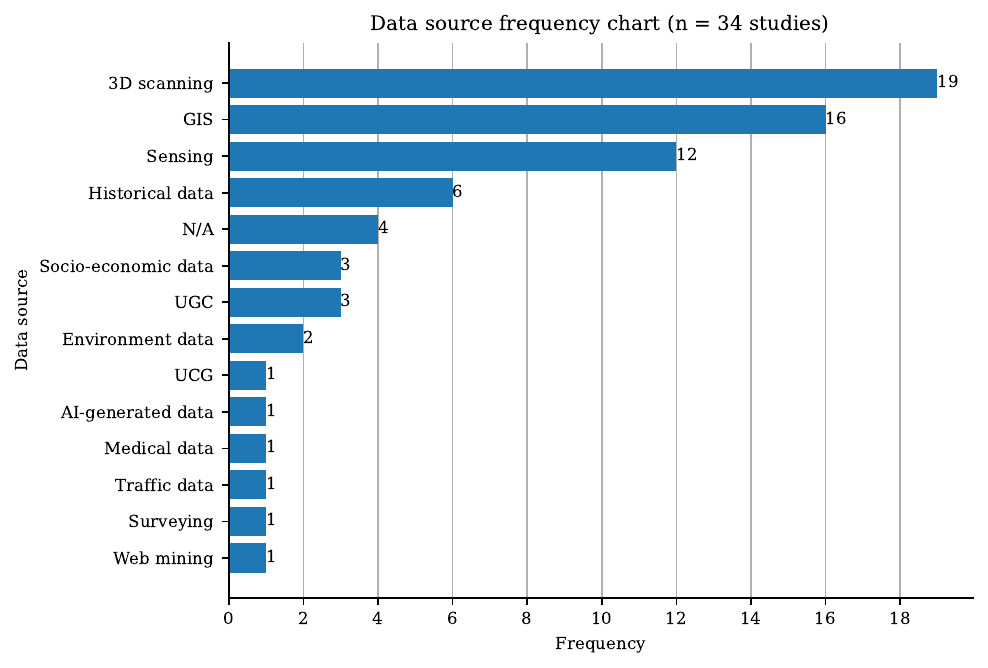}
    \caption{Frequency of data source categorizations in studies (Source: Authors own work)}
    \label{fig:data_sources_freq}
\end{figure}

Figure \ref{fig:data_sources_freq} depicts the frequency of studies categorized in each data source dimension.
The most used data source is 3D scanning. This source consists of multiple scanning technologies (e.g., LiDAR, photogrammetry, laser scanning), capturing the physical environment of the tourism site, allowing the creation of digital 3D models to be incorporated into the DTs. Its main disadvantages are the need for manual input and only capturing static data. Changes in the environment require subsequent human intervention. $N = 19$ studies use this technology to digitalize the physical environments of the sites. The data is complemented with Heritage Building Information Modeling (HBIM) to represent heritage buildings and provide accurate digital representations. 
Another extensively used data source is GIS data, used in  $N = 16$ studies. This georeferenced data can come from multiple providers (e.g., propriety data, open source data) and data gathered in-house, and could be served and updated frequently. This data often complements 3D scanning for the accurate relative positioning of scanned artifacts.
$N = 12$ studies use or propose using sensing in their presented DTs. This source provides data to the DT using Internet of Things (IoT) powered sensors, capturing dynamic data from the physical entity (e.g., tourist activity and counting, environmental data) that is hard to gather from other sources and enriching the DT. This data can be gathered and processed in real or near-real time and used by the DT system to adapt to the state of the real entity in adequate time. 
Historical data is used in some studies to feed the DT ($N = 6$) \citep{Zhang2022, Guzzetti2022, Banfi2022, Zhao2022, Chen2023, Feklistov2024}. This data consists of any type of data that records past circumstances of tourism objects of interest. Historical documents describing the object and art imagery are examples of this data. It is relevant to digitally restore past versions of the tourism object and capture art data. 
UCG is relevant, although neglected, as a data source for DT applications, as only $N = 3$ of the reviewed studies indicate the source \citep{Seungyoub2021, Sang2022, Prabawati2024}. As the name implies, the data is obtained from content generated and posted by users on online platforms. Examples include social media posts, blogs, tourism reviews, and surveys. This data is valuable for capturing other dynamic aspects of the physical entity, such as tourist satisfaction, site attractivity, and general tourist activity. 
Web mining also captures dynamic tourism data, similar to UGC (which it can also capture). The source pulls static or dynamic data about tourism objects from the web, mainly through web scraping or extracting data from Application Programming Interfaces (API) of target websites (e.g., X, former Twitter). Web-mined big data has enormous potential to feed tourism DTs and extract unknown knowledge about the tourism system. Only one study refers to using this source \citep{Sang2022}.
Environmental and socioeconomic data are data sources to feed DTs with information about the environment of a tourism point of interest and its socioeconomic circumstances, respectively. $N = 3$ studies \citep{Boletsis2022, Li2024, Kourtit2025} use this data.

Other less-used but worth mentioning data sources include medical data and traffic \citep{Li2024} to complement the DT with relevant information in these areas and AI-generated data \citep{Dayoub2024} for theoretical prototyping purposes, and surveying, obtaining data from people, typically experts or stakeholders, through surveys and questionnaires. \cite{Litavniece2023} use this source to gather knowledge of tourism product competitiveness for the proposed DT.
$N = 4$ studies do not disclose data sources used or suggested (classified as "N/A").

\subsubsection{Data linkage}
\begin{figure}[H]
    \centering
    \includegraphics[width=0.5\textwidth]{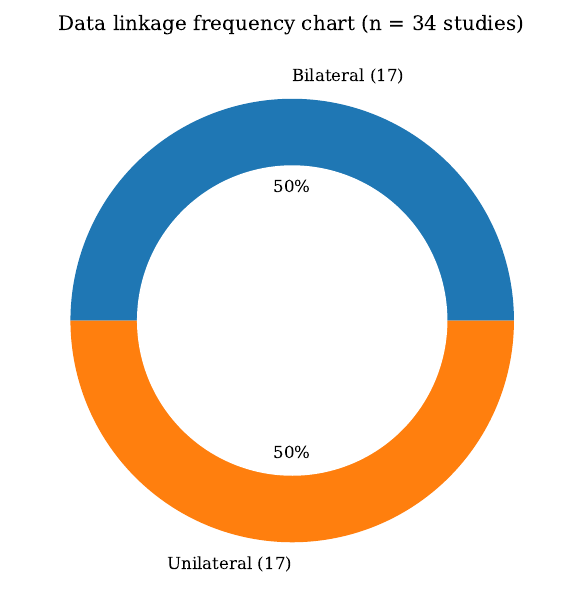}
    \caption{Frequency of data linkage categorization in studies (Source: Authors own work)}
    \label{fig:data_linkage_freq}
\end{figure}

Figure \ref{fig:data_linkage_freq} depicts the frequency of studies categorized in each data linkage dimension.
Data linkage refers to the directionality of the data flows between twins. Unilateral data flow means only the digital counterpart receives data about the physical entity. Bilateral flow means that the data comes not only from physical to digital but also from digital to physical in the form of data analysis, prediction, feedback, and, in more advanced cases, complete control (which may not be adequate for many tourism applications, e.g., tourism services which require human control or are traditionally hard to automate) to alter the physical entity. Unilateral data flow does not allow the DTs to provide this feedback. 
Most of the main definitions of DT acknowledge a true DT as bilateral \citep{vanderValk2022}, as "twinning" implies a strong bond between physical and digital, as they must replicate each other. There is an equal amount of studies ($N1, N2 = 17$) that use unilateral data flow, meaning an incomplete twinning, and bilateral data connection, usually in indirect influence (i.e., there is no automated control of the physical process, such as automatically redirecting tourism flow. The changes in the physical entity come from decision-makers when acting upon the data generated by the DT), which we consider in this work. These studies did not have a technical, "true" twinning connection, where the physical entity changes to match the results of the digital counterpart derived from data flow and processing of simulations.
This suggests that tourism DT studies often use a simplistic definition of twinning, such as "representing a physical environment in digital means" or "creating a digital representation of a physical entity", not acknowledging the bilateral connection between twins present in its main definitions.  

\subsubsection{Visualization}

\begin{figure}[H]
    \centering
    \includegraphics[width=0.75\textwidth]{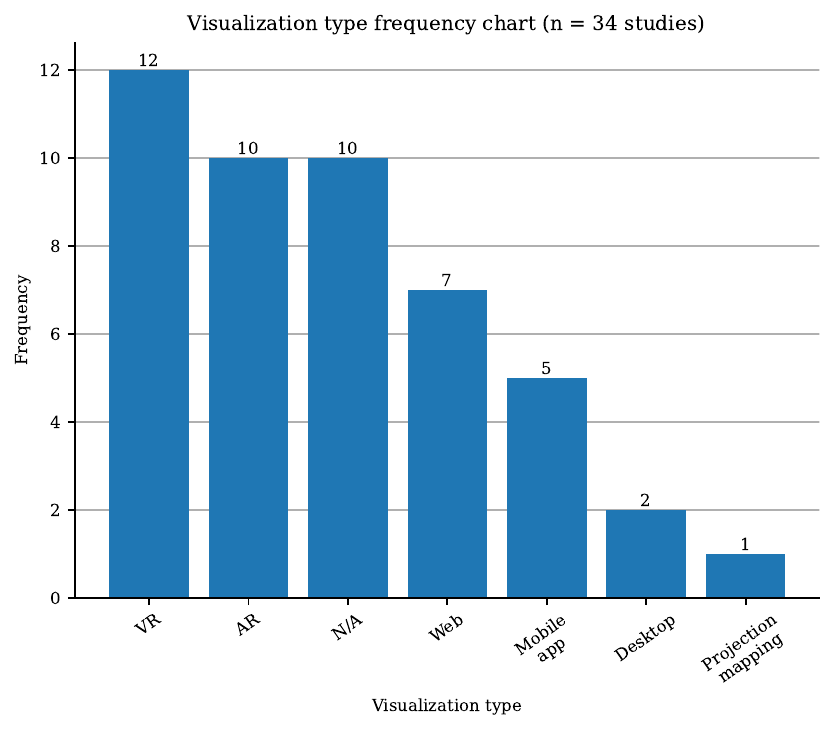}
    \caption{Frequency of visualization categorization in studies (Source: Authors own work)}
    \label{fig:visualization_freq}
\end{figure}

Figure \ref{fig:visualization_freq} depicts the frequency of studies categorized in each visualization dimension.
Extended Reality (XR) technologies, such as VR and AR, are greatly used as visualization and Human-Machine Interface (HMI) technologies in the DTs proposed in the reviewed studies. VR technology allows the immersive exploration of digitalized tourist objects by simulating physical worlds in the digital space, usually with specialized goggles, to fully immerse into the digitalized world in the comfort of the cyber-tourists' homes. AR facilitates enhancing the physical environment through mobile phones or XR goggles equipped with cameras that capture the actual space and add digitalized elements into the physical entity through computer graphics, incrementing basic explorations of the scenic spots with often hidden or historical information. VR is used or suggested in $N = 12$, and AR (separately or together) is used in $N = 10$ studies.

Some studies propose or use mobile apps ($N = 5$) and simple (without XR) web applications ($N = 7$). Since smart devices and internet connections are widely available to users worldwide today, this is an accessible way to reach wider audiences and beat VR and AR applications that require special goggles regarding entry barriers to using the DT if intended for public-wide use.
Mobile apps and web apps are usually proposed together in the same studies to complement each other \citep{Zhang2023, Sang2022, Planu2024, Litavniece2023, Prabawati2024}, allowing the users to choose the medium they prefer.

\cite{Boletsis2022} uses the innovative projection mapping interfacing technology to enrich a 3D scaled physical model of the geology of the scenic spot with relevant data visualized through the projection upon the model.  
\cite{Galiano2024} use a simple desktop application to interact with the DT.
\cite{Prabawati2024} present a framework to implement DTs for tourism, proposing almost all aforementioned technologies when suitable: web, mobile, desktop and VR.
The studies classified as "N/A" do not specify the visualization and HMI used to interact with the DT.

\subsubsection{Application}

\begin{figure}[H]
    \centering
    \includegraphics[width=0.5\textwidth]{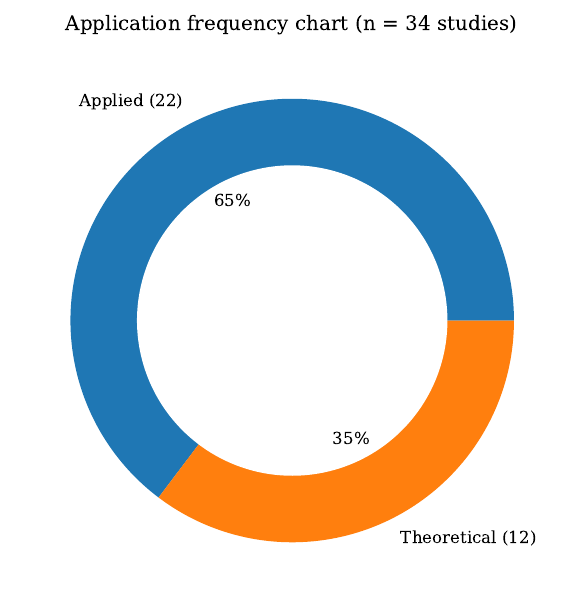}
    \caption{Frequency of studies per application dimension (Source: Authors own work)}
    \label{fig:application_freq}
\end{figure}

Figure \ref{fig:application_freq} shows the frequency of studies categorized in each application dimension. We can observe most studies ($N = 22$) present DT implemented in real-life scenarios or applied prototypes. The remaining studies ($N = 12$) present theoretical and conceptual frameworks or guidelines for creating tourism DTs or aiding in one or more processes of creating DTs backed by theoretical concepts and studies.
Applied studies generally have more impact than theoretical studies, as the former often have a more solid foundation derived from empirical evidence, and the benefits of their usage are better observed. However, theoretical studies are also important in paving the way for applied studies, as they guide the implementation of applied studies with theory.
The abundance of applied studies in the reviewed studies indicates the growing interest in real applications of this technology.

\subsubsection{Summary}
In Table \ref{tab:summary}, the classification of each reviewed study is presented and summarized. Classifications marked with "-" represent "N/A". In the next subsection, the results are discussed.

\begin{landscape}
\renewcommand{\arraystretch}{0.3} 
\begin{table}[H]
\centering
\caption{Review summary table (Source: Authors own work)}
\label{tab:summary}
\small
\resizebox{\linewidth}{0.26\paperheight}{%
\begin{tabular}{p{7cm}lllllll}
\toprule
Study & Tourism type & Application & Data source & Data linkage & Visualization type & Spatial scale & Purpose\\[0.1cm]
\midrule
\cite{LopezGonzalez2023} & C, U & A & 3DS, GIS, S & B & AR & S & EE \\[0.2cm]
\cite{Rahmadian2023} & G & T & - & B & - & R & DM \\[0.2cm]
\cite{Gallist2023} & U & A & 3DS & U & VR & L, S & VT \\[0.2cm]
\cite{Zhou2023} & C & A & 3DS, GIS & U & - & L & DM, EE, VT \\[0.2cm]
\cite{Zhang2023} & G & T & GIS, S & U & MA, W & L, S & DM, EE \\[0.2cm]
\cite{Kumar2022} & C & A & 3DS, S & U & VR & S & DM \\[0.2cm]
\cite{Yang2021} & C, U & T & - & B & - & L & DM \\[0.2cm]
\cite{Seungyoub2021} & S & T & S, UGC & B & AR, VR & L & DM \\[0.2cm]
\cite{Zhang2022} & C & A & 3DS, HD & U & AR, VR & S & DM, EE, VT \\[0.2cm]
\cite{Guzzetti2022} & C & A & 3DS, GIS, HD & U & VR & S & DM, EE, VT \\[0.2cm]
\cite{Rahmadian2023Governing} & G & T & - & B & - & R & DM \\[0.2cm]
\cite{Sang2022} & C, R & A & 3DS, GIS, S, UGC, WM & B & MA, W & L & DM, VT \\[0.2cm]
\cite{Banfi2022} & C & A & 3DS, HD & U & AR, VR & S & DM, EE, VT \\[0.2cm]
\cite{Zhao2022} & C & T & HD, S & U & AR, VR & S & DP, VT \\[0.2cm]
\cite{Boletsis2022} & N & A & ED, SED & B & PM & L & VT \\[0.2cm]
\cite{FischerStabel2021} & N & A & 3DS, GIS & U & AR & S & EE \\[0.2cm]
\cite{Litavniece2023} & G & T & GIS, Su & B & MA, W & L & DM \\[0.2cm]
\cite{Logothetis2023} & C, N & A & 3DS, GIS & U & AR & S & DM, VT \\[0.2cm]
\cite{Nguyen2022} & C & A & 3DS, GIS & U & AR, VR & S & DM, EE, VT \\[0.2cm]
\cite{Doria2023} & C & A & 3DS, GIS & U & - & S & DM, DP \\[0.2cm]
\cite{Li2024} & U & A & MD, SED, TD & B & - & L & DM \\[0.2cm]
\cite{Dayoub2024} & C & T & AIGD & U & - & L, S & VT \\[0.2cm]
\cite{Fistola2023} & C, U & A & 3DS & U & AR & L, S & EE \\[0.2cm]
\cite{Rolph2024} & C, N & A & ED, GIS & B & W & L, R & DM \\[0.2cm]
\cite{Galiano2024} & C & A & 3DS, S & U & D & S & DM \\[0.2cm]
\cite{Chen2023} & C & T & 3DS, HD, S, UGC & B & - & L & DM, VT \\[0.2cm]
\cite{Tang2024} & C & A & 3DS, GIS & B & AR, VR & S & DM, VT \\[0.2cm]
\cite{Deng2024} & G & T & - & U & - & - & VT \\[0.2cm]
\cite{Kourtit2025} & C, U & A & GIS, SED & B & - & L & DM \\[0.2cm]
\cite{Chen2024} & C & A & 3DS, S & B & W & S & DM, VT \\[0.2cm]
\cite{Feklistov2024} & C & A & GIS, HD & U & VR & L, S & VT \\[0.2cm]
\cite{Huang2024} & C & A & S & B & VR & L, S & DM, VT \\[0.2cm]
\cite{Planu2024} & C, Ga, N & T & 3DS, GIS, S & B & MA, W & L & DM \\[0.2cm]
\cite{Prabawati2024} & G & T & 3DS, GIS, S, UCG & B & D, MA, VR, W & S & DM, VT \\[0.2cm]
\bottomrule
\end{tabular}
}%
\end{table}
\end{landscape}

\subsection{Discussion}

For each of the proposed research questions, there are our following discussions:

\subsubsection{RQ1: How developed is the state-of-the-art of digital twinning in tourism?}

The related studies on DT in tourism are very recent, starting in 2020. So, very few studies present the technology's applied or theoretical application in tourism-related applications. Although the topic is starting to gain traction, it is showing fertile ground for its evolution.
Europe and Asia are the continents publishing peer-reviewed papers about concrete applications of DT technology in tourism in scientific journals, conferences, and books. China has the biggest author-wise contribution to these studies. Their contributions are numerically similar. Other continents, specifically countries, may also contribute to tourism DT, seemingly outside the academic environment.

Many of the studies found are conceptual and theoretical, proposing frameworks to create the DTs, which have not yet been applied in real-life situations. Most of them have been implemented and tested in real-life tourism applications.
According to some definitions of DT, there should be an automatic data linkage between the physical and digital twins. Many of the reviewed studies have a unilateral data linkage with the digital counterpart, which, according to said definitions, may not be considered "authentic" DTs. The ones with bilateral data have an indirect digital-to-physical link, where the physical entity changes from the feedback provided by the DT through external means.

\subsubsection{RQ2: How are digital twins used for tourism-related applications?}

We found applied or theoretical DTs, mostly in cultural heritage tourism. Other studies were focused on natural, urban, rural, and science tourism, and some could be applied in various tourism types, being considered generic.
There is a heavy focus on the digitalization of touristic scenic spots. 3D scanning and GIS data are the leading technologies used in these artifacts. They are used mainly to integrate VR visitation in these places when physical visitation is impractical or to enhance in-situ visitation tourism sites using AR. In these scenarios, the concept of DT ignores the bilateral communication between the original and digital twins, with only the DT receiving data from the original. These DTs only simulate or enhance visitation without capturing dynamic tourism system data. Some studies harness dynamic data sources, such as sensing technologies, UGC mining, and web mining, allowing data collection about tourist behavior, monitoring tourism activity, and predicting crowds. These DTs will enable the monitoring and possibly predicting how tourists behave, providing feedback to stakeholders to change how the services are implemented. The primary purposes of the DT systems for tourism are the digital preservation of heritage sites, destination management, virtual tourism, and the enhancement of physical tourism experiences. The spatial scale of the reviewed DT systems spans from site-level to regional level, with the vast majority applying the former scale, possibly due to general interest in managing individual tourism sites and the great complexity required in larger-scale systems.

\subsubsection{RQ3: What are the research gaps and future directions of digital twins for tourism?}

DT in tourism is still at an early stage of development and adherence in the academic scene. There is an emphasis on using DT for destination management and virtual tourism. There is a gap in research on putting DT technology into practice to enhance the tourism experience, although some studies already use DT for this purpose.
Few types of tourism benefit from DT technology. Six specific types of tourism were extracted from the literature. The remaining studies can be applied to different types of tourism but are theoretical. This leaves an opportunity to test and use the technology to other types of tourism, which could benefit equally.
The data sources used to feed digital counterparts and give feedback to the physical entity are limited. Although data types are adequately used to create the digital version of the twin, the literature's heavy focus on static data sources and unilateral data linkage leads to the absence of quality feedback to improve tourism activity.
As tourists are a crucial part of tourism, monitoring their activities is an essential part of DT-based systems for tourism. Sources for this dynamic data, such as sensing, UCG, and web mining, should be used for a complete DT, enabling the prediction and monitoring of over-tourism situations and tourism service quality given real or fabricated tourism activity scenarios.
Using bilateral data linkage and dynamic data sources, DTs can provide more benefits, becoming "more real," according to its main definitions \citep{vanderValk2022}, having more impact on their physical counterparts, allowing the prediction of and prescription for tourism futures. Tourism-applied DTs generally lack this full-duplex data connection; as such, creating bilaterally linked DTs for tourism is a future direction to follow.
A possible future direction for tourism regulation agencies is applying DT in wider-scoped scenarios, such as city-wide or country-wide tourism. The current literature focuses on applications of site-level and local-level spatial scales (with a few regional-level DTs). As such, there is an opportunity to explore the application of DT technology to more macroscopic scales, representing regional, national, or global tourism systems. Important data sources include tourism census data, such as accommodation numbers, border control data, and cell phone data.

\textbf{For researchers:}
DTs are valuable tools for scholars. The digital representation of tourism systems, backed by data gathering, allows for in-silico experiments on hard-to-observe scenarios of a tourism system with academic interest. Higher spatial scale tourism systems may be more attractive to researchers than to practitioners, as the latter typically operate at the site or local level, where their business operations run. Regional and above scopes can be more helpful to researchers or high-level/governmental stakeholders to understand macroscopic, large-scale tourism dynamics. Researchers can also work with practitioners on low-level scales to scientifically improve and gain insight into microscopic tourism systems.

\textbf{For practitioners:} The development of DT systems may also help practitioners manage their destinations and improve service quality altogether. The monitoring, predicting, and prescribing qualities of a DT, backed by the static and dynamic data collected from the physical counterpart, allow for improving the destination management operations in a safe environment, discovering service faults or bottlenecks that could be improved, and providing service alternatives. Advanced ideal solutions would even allow the real-life tourism service to automatically adapt to the evolution of the digital counterpart, being a true DT application and reducing direct practitioner involvement.

\subsubsection{Validity threats}

Construct validity threat relates to whether the selected studies and criteria adequately reflect the concepts we aim to explore (e.g., the application of Digital Twin (DT) technology in tourism). The reliance on three scientific databases (Scopus, Web of Science, and Dimensions) may exclude relevant studies not indexed in these repositories, potentially leading to incomplete conceptual coverage. Further exploitation of snowballing and including grey literature could enrich this review.

Internal validity concerns potential biases in the selection and classification of studies. For instance, applying inclusion and exclusion criteria might inadvertently favour specific study types or methodologies. We mitigated this by following a systematic review methodology, as outlined in \cite{CarreraRivera2022}, and employing a team-based approach to discuss and define the selection and classification protocol and to screen and categorize studies.

We cannot claim the findings' generalizability (external validity threat), but we believe this SLR provides a good picture of the current state of the art. The limited number of studies (34) reviewed is, in fact, due to the emerging nature of DT applications in tourism.

\section{Conclusions}
\label{Conclusions&FutureWork}

We presented an SLR to assess the current state of the art in applying DT technology in tourism-related applications. The methodology used to conduct the review was detailed to allow for replication. A total of 34 peer-reviewed studies from scientific databases were selected for review. A bibliometric analysis was performed to understand the demographics of the studies. A taxonomy of seven criteria was used to classify the studies: tourism type, purpose, spatial scale, data sources, data linkage, visualization, and application. 

The main conclusions drawn from the review are the following:
Europe and Asia are the continents publishing scientific studies on the application of DT technology in tourism with similar contributions, with China being the most active country in this regard. The topic is at an early stage of adoption (the ``oldest'' study found is from 2021) but is gaining interest every year.

Most of the reviewed studies focus on cultural tourism and heritage digitalization. Six other types of tourism were detected. There is an emphasis on static data sources, e.g., to generate digital counterpart models. Some studies have dynamic data sources that capture tourism activity, but their usage is often neglected. The data linkage between real and digital counterparts is mainly unilateral, so the real entity cannot evolve or synchronize with the results of the DT processing, failing to comply with the usual DT requirements for the bidirectional connection.

Most DT studies applied to tourism benefit from XR technologies, which allow users to interact with and visualize digital representations. Other simpler or more innovative interfaces were used. There are more applied studies than theoretical ones, suggesting an advance in applying the technology in the field.
Finally, we conclude that there is a vast research gap regarding DT technology in tourism that is worth filling.

\section*{Acknowledgments}

\blindreview{This work was partially funded by the \href{https://www.fct.pt/en/}{Portuguese Foundation for Science and Technology (FCT)}, under \href{https://istar.iscte-iul.pt/}{ISTAR-Iscte} project UIDB/04466/2020.}


\bibliographystyle{agsm}
\bibliography{main, slr}

@inproceedings{Abreu2024,
	title        = {{A digital transformation approach to scaffold tourism crowding management: pre-factum, on-factum, and post-factum}},
	author       = {{Brito e Abreu}, Fernando and {Neto Marinheiro}, Rui and {Boavida-Portugal}, Inês and Lopes, Adriano and {Mestre Santos}, Tomás and {Sampaio de Almeida}, Duarte and Simões, Rodrigo},
	year         = 2024,
	month        = {Jan.~31--Feb.~3},
	booktitle    = {Proc. of the 9th International Conference on Digital Arts, Media and Technology and 7th ECTI Northern Section Conference on Electrical, Electronics, Computer and Telecommunications Engineering (ECTI DAMT \& NCON 2024)},
	address      = {Chiangmai, Thailand},
	pages        = {586--591},
	doi          = {10.1109/ECTIDAMTNCON60518.2024.10480056},
	isbn         = {979-8-3503-1824-1},
	issn         = {2768-4644},
	organization = {IEEE}
}

@techreport{AIAA2020,
	title        = {{Digital Twin: Definition \& Value -- An AIAA and AIA Position Paper}},
	author       = {{AIAA Digital Engineering Integration Committee}},
	year         = 2020,
	address      = {Reston, VA, USA},
	url          = {https://www.aia-aerospace.org/wp-content/uploads/Digital-Twin-Institute-Position-Paper-December-2020-1.pdf},
	institution  = {American Institute of Aeronautics and Astronautics (AIAA) \& Aerospace Industries Association (AIA)}
}

@article{Arthur2020,
	title        = {{Digital Twin: Definition \& Value—AIAA and AIA Position Paper}},
	author       = {Arthur, R and French, M and Ganguli, J and Kinard, DA and Kraft, E and Marks, I and Matlik, J and Fischer, O and Sangid, M and Seal, D and others},
	year         = 2020,
	journal      = {AIAA Digital Engineering Integration Committee}
}

@article{Barricelli2019,
	author       = {Barricelli, Barbara Rita and Casiraghi, Elena and Fogli, Daniela},
	journal      = {IEEE Access},
	title        = {{A Survey on Digital Twin: Definitions, Characteristics, Applications, and Design Implications}},
	year         = 2019,
	volume       = 7,
	number       = {},
	pages        = {167653--167671},
	doi          = {10.1109/ACCESS.2019.2953499}
}

@article{Bekele2024,
	title        = {{Digitalization and digital transformation in the tourism industry: a bibliometric review and research agenda}},
	author       = {Bekele, Henok and Raj, Sahil},
	journal      = {Tourism Review},
	year         = 2024,
	publisher    = {Emerald Publishing Limited}
}

@article{BotinSanabria2022,
	title        = {{Digital Twin Technology Challenges and Applications: A Comprehensive Review}},
	author       = {Botín-Sanabria, Diego M. and Mihaita, Adriana-Simona and Peimbert-García, Rodrigo E. and Ramírez-Moreno, Mauricio A. and Ramírez-Mendoza, Ricardo A. and Lozoya-Santos, Jorge de J.},
	year         = 2022,
	journal      = {Remote Sensing},
	volume       = 14,
	number       = 6,
	doi          = {10.3390/rs14061335},
	issn         = {2072-4292},
	article-number = 1335
}

@article{Buhalis2024,
	title        = {{Metaverse as a disruptive technology revolutionising tourism management and marketing}},
	author       = {Dimitrios Buhalis and Daniel Leung and Michael Lin},
	year         = 2023,
	journal      = {Tourism Management},
	volume       = 97,
	pages        = 104724,
	doi          = {https://doi.org/10.1016/j.tourman.2023.104724},
	issn         = {0261-5177}
}

@article{CarreraRivera2022,
	title        = {{How-to conduct a systematic literature review: A quick guide for computer science research}},
	author       = {Angela Carrera-Rivera and William Ochoa and Felix Larrinaga and Ganix Lasa},
	year         = 2022,
	journal      = {MethodsX},
	volume       = 9,
	pages        = 101895,
	doi          = {10.1016/j.mex.2022.101895},
	issn         = {2215-0161}
}

@article{Casal2023,
	title        = {{Review of crisis management frameworks in tourism and hospitality: A meta-analysis approach}},
	author       = {Casal-Ribeiro, Mariana and Boavida-Portugal, In{\^e}s and Peres, Rita and Seabra, Cl{\'a}udia},
	year         = 2023,
	journal      = {Sustainability},
	publisher    = {MDPI},
	volume       = 15,
	number       = 15,
	pages        = 12047
}

@inproceedings{Glaessgen2012,
	title        = {{The digital twin paradigm for future NASA and US Air Force vehicles}},
	author       = {Glaessgen, Edward and Stargel, David},
	year         = 2012,
	booktitle    = {Proc. of the 53rd Structures, Structural Dynamics, and Materials Conference},
	pages        = 1818,
	doi          = {10.2514/6.2012-1818}
}

@article{Grieves2005,
	author       = {Grieves, Michael},
	year         = 2005,
	month        = {01},
	pages        = {},
	title        = {{Product lifecycle management: the new paradigm for enterprises}},
	volume       = 2,
	journal      = {International Journal of Product Development - Int J Prod Dev},
	doi          = {10.1504/IJPD.2005.006669}
}

@article{Grieves2014,
	title        = {{Digital twin: manufacturing excellence through virtual factory replication}},
	author       = {Grieves, Michael},
	year         = 2014,
	journal      = {White paper},
	volume       = 1,
	number       = 2014,
	pages        = {1--7}
}

@incollection{Grieves2017,
	title        = {{Digital Twin: Mitigating Unpredictable, Undesirable Emergent Behavior in Complex Systems}},
	author       = {Grieves, Michael and Vickers, John},
	year         = 2017,
	booktitle    = {Transdisciplinary Perspectives on Complex Systems: New Findings and Approaches},
	publisher    = {Springer International Publishing},
	address      = {Cham},
	pages        = {85--113},
	doi          = {10.1007/978-3-319-38756-7_4},
	isbn         = {978-3-319-38756-7},
	editor       = {Kahlen, Franz-Josef and Flumerfelt, Shannon and Alves, Anabela}
}

@article{Haag2019,
	title        = {{Automated Generation of as-manufactured geometric Representations for Digital Twins using STEP}},
	author       = {Sebastian Haag and Reiner Anderl},
	year         = 2019,
	journal      = {Procedia CIRP},
	volume       = 84,
	pages        = {1082--1087},
	doi          = {10.1016/j.procir.2019.04.305},
	issn         = {2212-8271},
	note         = {29th CIRP Design Conference 2019, 08-10 May 2019, Póvoa de Varzim, Portgal}
}

@inproceedings{Heluany2023,
	title        = {{Survey on Digital Twins: from concepts to applications}},
	author       = {B. Heluany, Jessica and Gkioulos, Vasileios},
	year         = 2023,
	booktitle    = {Proc. of the 18th International Conference on Availability, Reliability and Security (ARES'23)},
	location     = {, Benevento, Italy,},
	publisher    = {Association for Computing Machinery},
	address      = {New York, NY, USA},
	doi          = {10.1145/3600160.3605070},
	isbn         = 9798400707728,
	articleno    = 119,
	numpages     = 8
}

@article{Hu2021,
	author       = {Hu, Weifei and Zhang, Tongzhou and Deng, Xiaoyu and Liu, Zhenyu and Tan, Jianrong},
	title        = {{Digital twin: a state-of-the-art review of its enabling technologies, applications and challenges}},
	journal      = {Journal of Intelligent Manufacturing and Special Equipment},
	year         = 2021,
	month        = {Jan},
	day          = {01},
	publisher    = {Emerald Publishing Limited},
	volume       = 2,
	number       = 1,
	pages        = {1--34},
	doi          = {10.1108/JIMSE-12-2020-010}
}

@article{Jones2020,
	title        = {{Characterising the Digital Twin: A systematic literature review}},
	author       = {David Jones and Chris Snider and Aydin Nassehi and Jason Yon and Ben Hicks},
	year         = 2020,
	journal      = {CIRP Journal of Manufacturing Science and Technology},
	volume       = 29,
	pages        = {36--52},
	doi          = {10.1016/j.cirpj.2020.02.002},
	issn         = {1755-5817}
}

@article{PirilloRamos2021,
	title        = {{Tourism-phobia in Barcelona: dismantling discursive strategies and power games in the construction of a sustainable tourist city}},
	author       = {Silvana Pirillo Ramos and Lluis Mundet},
	year         = 2021,
	journal      = {Journal of Tourism and Cultural Change},
	publisher    = {Routledge},
	volume       = 19,
	number       = 1,
	pages        = {113--131},
	doi          = {10.1080/14766825.2020.1752224}
}

@article{Prisma2020,
	title        = {{The PRISMA 2020 statement: an updated guideline for reporting systematic reviews}},
	author       = {Page, Matthew J and McKenzie, Joanne E and Bossuyt, Patrick M and Boutron, Isabelle and Hoffmann, Tammy C and Mulrow, Cynthia D and Shamseer, Larissa and Tetzlaff, Jennifer M and Akl, Elie A and Brennan, Sue E and others},
	year         = 2021,
	journal      = {Bmj},
	publisher    = {British Medical Journal Publishing Group},
	volume       = 372
}

@article{Rosen2015,
	title        = {{About The Importance of Autonomy and Digital Twins for the Future of Manufacturing}},
	author       = {Roland Rosen and Georg {von Wichert} and George Lo and Kurt D. Bettenhausen},
	year         = 2015,
	journal      = {IFAC-PapersOnLine},
	volume       = 48,
	number       = 3,
	pages        = {567--572},
	doi          = {10.1016/j.ifacol.2015.06.141},
	issn         = {2405-8963},
	note         = {15th IFAC Symposium on Information Control Problems in Manufacturing}
}

@article{Semeraro2021,
	title        = {{Digital twin paradigm: A systematic literature review}},
	journal      = {Computers in Industry},
	volume       = 130,
	pages        = 103469,
	year         = 2021,
	issn         = {0166-3615},
	doi          = {10.1016/j.compind.2021.103469},
	author       = {Concetta Semeraro and Mario Lezoche and Hervé Panetto and Michele Dassisti}
}

@article{Tao2019,
	title        = {{Digital twin-driven product design framework}},
	author       = {Fei Tao and Fangyuan Sui and Ang Liu and Qinglin Qi and Meng Zhang and Boyang Song and Zirong Guo and Stephen C.-Y. Lu and A. Y. C. Nee},
	year         = 2019,
	journal      = {International Journal of Production Research},
	publisher    = {Taylor \& Francis},
	volume       = 57,
	number       = 12,
	pages        = {3935--3953},
	doi          = {10.1080/00207543.2018.1443229}
}

@book{UNWTO2018,
	title        = {{‘Overtourism'? – Understanding and Managing Urban Tourism Growth beyond Perceptions, Executive Summary}},
	author       = {{World Tourism Organization (UNTWO)} and {Centre of Expertise Leisure Tourism \& Hospitality Sciences} and {NHTV Breda University of Applied Sciences} and {NHL Stenden University of Applied Sciences}},
	year         = 2018,
	booktitle    = {‘Overtourism'? – Understanding and Managing Urban Tourism Growth beyond Perceptions, Executive Summary},
	publisher    = {World Tourism Organization (UNWTO)},
	doi          = {10.18111/9789284420070}
}

@article{vanderValk2022,
	title        = {{Archetypes of Digital Twins}},
	author       = {van der Valk, Hendrik and Ha{\ss}e, Hendrik and M{\"o}ller, Frederik and Otto, Boris},
	year         = 2022,
	month        = {Jun},
	day          = {01},
	journal      = {Business \& Information Systems Engineering},
	volume       = 64,
	number       = 3,
	pages        = {375--391},
	doi          = {10.1007/s12599-021-00727-7},
	issn         = {1867-0202}
}

@article{Verma2022,
	title        = {{Past, present, and future of virtual tourism-a literature review}},
	author       = {Sanjeev Verma and Lekha Warrier and Brajesh Bolia and Shraddha Mehta},
	year         = 2022,
	journal      = {International Journal of Information Management Data Insights},
	volume       = 2,
	number       = 2,
	pages        = 100085,
	doi          = {10.1016/j.jjimei.2022.100085},
	issn         = {2667-0968}
}

@article{Wagner2019,
	title        = {{Challenges and Potentials of Digital Twins and Industry 4.0 in Product Design and Production for High Performance Products}},
	author       = {Raphael Wagner and Benjamin Schleich and Benjamin Haefner and Andreas Kuhnle and Sandro Wartzack and Gisela Lanza},
	year         = 2019,
	journal      = {Procedia CIRP},
	volume       = 84,
	pages        = {88--93},
	doi          = {10.1016/j.procir.2019.04.219},
	issn         = {2212-8271},
	note         = {29th CIRP Design Conference 2019, 08-10 May 2019, Póvoa de Varzim, Portgal}
}

@article{Banfi2022,
	title        = {{Building archaeology informative modelling turned into 3D volume stratigraphy and extended reality time-lapse communication}},
	author       = {Banfi, Fabrizio and Brumana, Raffaella and Landi, Angelo Giuseppe and Previtali, Mattia and Roncoroni, Fabio and Stanga, Chiara and others},
	year         = 2022,
	journal      = {Virtual Archaeology Review},
	volume       = 13,
	number       = 26,
	pages        = {1--21},
	doi          = {10.4995/var.2022.15313}
}

@inproceedings{Boletsis2022,
	title        = {{The Gaia System: A Tabletop Projection Mapping System for Raising Environmental Awareness in Islands and Coastal Areas}},
	author       = {Boletsis, Costas},
	year         = 2022,
	booktitle    = {Proc. of the 15th International Conference on PErvasive Technologies Related to Assistive Environments (PETRA)},
	location     = {Corfu, Greece},
	publisher    = {Association for Computing Machinery},
	address      = {New York, NY, USA},
	pages        = {50–54},
	doi          = {10.1145/3529190.3535340},
	isbn         = 9781450396318
}

@inproceedings{Chen2023,
	title        = {{The Framework of Culture Heritage Digital Twin System Based on Semantic Network}},
	author       = {Chen, Liling and Song, Yicong and Qin, Shengfeng and Niu, Xiaojing and Yang, Liu and Luan, Xin},
	year         = 2023,
	booktitle    = {Proc. of the Smart World Congress (SWC)},
	pages        = {1--6},
	doi          = {10.1109/SWC57546.2023.10449294},
	organization = {IEEE}
}

@article{Chen2024,
	title        = {{Application of Remote Sensing and Computing in Smart Archaeology and Tourism—A Loulan Perspective}},
	author       = {Chen, Danling},
	year         = 2024,
	journal      = {Tourism Management and Technology Economy},
	volume       = 7,
	number       = 2,
	pages        = {88--97},
	doi          = {10.23977/tmte.2024.070213},
	doi          = {10.23977/tmte.2024.070213},
	note         = {https://doi.org/10.23977/tmte.2024.070213}
}

@article{Dang2023,
	title        = {{Digital twin applications on cultural world heritage sites in China: A state-of-the-art overview}},
	author       = {Xinyuan Dang and Wanqin Liu and Qingyuan Hong and Yibo Wang and Xuemin Chen},
	year         = 2023,
	journal      = {Journal of Cultural Heritage},
	volume       = 64,
	pages        = {228--243},
	doi          = {10.1016/j.culher.2023.10.005},
	issn         = {1296-2074}
}

@article{Dayoub2024,
	title        = {{Digital Silk Roads: Leveraging the Metaverse for Cultural Tourism within the Belt and Road Initiative Framework}},
	author       = {Dayoub, Bashar and Yang, Peifeng and Omran, Sarah and Zhang, Qiuyi and Dayoub, Alaa},
	year         = 2024,
	journal      = {Electronics},
	publisher    = {MDPI},
	volume       = 13,
	number       = 12,
	pages        = 2306,
	doi          = {10.3390/electronics13122306}
}

@article{Deng2024,
	title        = {{From metaverse experience to physical travel: the role of the digital twin in metaverse design}},
	author       = {Deng, Baolin and Wong, IpKin Anthony and Lian, Qi Lilith},
	year         = 2024,
	journal      = {Tourism Review},
	publisher    = {Emerald Publishing Limited},
	doi          = {10.1108/TR-05-2023-0315}
}

@article{Doria2023,
	title        = {{Documentation, conservation, and reuse planning activities for disused cultural heritage}},
	author       = {Doria, Elisabetta and Morandotti, Marco},
	year         = 2023,
	month        = {Apr.},
	journal      = {VITRUVIO - International Journal of Architectural Technology and Sustainability},
	volume       = 8,
	pages        = {30–47},
	doi          = {10.4995/vitruvio-ijats.2023.18814}
}

@inproceedings{Feklistov2024,
	title        = {{Virtual Reality Tools for Creating Interactive Digital Twins of Attractions Infrastructure}},
	author       = {Feklistov, Vladislav and Gurtyakov, Alexander and Shuklin, Aleksey and Savina, Oksana and Ereshchenko, Tatyana},
	year         = 2024,
	booktitle    = {Proc. of the 4th International Conference on Novel \& Intelligent Digital Systems (NiDS)},
	publisher    = {Springer Nature Switzerland},
	address      = {Cham},
	pages        = {257--268},
	doi          = {10.1007/978-3-031-73344-4_21},
	isbn         = {978-3-031-73344-4},
	editor       = {Mylonas, Phivos and Kardaras, Dimitris and Caro, Jaime}
}

@incollection{FischerStabel2021,
	title        = {{Digital Twins, Augmented Reality and Explorer Maps Rising Attractiveness of Rural Regions for Outdoor Tourism}},
	author       = {Fischer-Stabel, Peter and Mai, Franziska and Schindler, Sabine and Schneider, Matthias},
	year         = 2021,
	booktitle    = {Advances and New Trends in Environmental Informatics},
	publisher    = {Springer International Publishing},
	address      = {Cham},
	pages        = {243--253},
	doi          = {10.1007/978-3-030-61969-5_17},
	isbn         = {978-3-030-61969-5},
	editor       = {Kamilaris, Andreas and Wohlgemuth, Volker and Karatzas, Kostas and Athanasiadis, Ioannis N.}
}

@inproceedings{Fistola2023,
	title        = {{Beyond the Smart City. The Urban Digital Twin for the Augmented City: The Vox Hortus Project}},
	author       = {Fistola, Romano and Zingariello, Ida},
	year         = 2023,
	booktitle    = {Proc. of the International Conference on Innovation in Urban and Regional Planning},
	pages        = {204--210},
	doi          = {10.1007/978-3-031-54118-6_19},
	organization = {Springer}
}

@article{Galiano2024,
	title        = {{The Influence of Visitors on Heritage Conservation: The Case of the Church of San Juan del Hospital, Valencia, Spain}},
	author       = {Galiano-Garrig{\'o}s, Antonio and L{\'o}pez-Gonz{\'a}lez, Concepci{\'o}n and Garc{\'\i}a-Valldecabres, Jorge and P{\'e}rez-Carrami{\~n}ana, Carlos and Emmitt, Stephen},
	year         = 2024,
	journal      = {Applied Sciences},
	booktitle    = {2023 IEEE Smart World Congress (SWC)},
	publisher    = {MDPI},
	volume       = 14,
	number       = 5,
	pages        = 2065,
	doi          = {10.3390/app14052065},
	organization = {IEEE}
}

@inproceedings{Gallist2023,
	title        = {{Tourism in the Metaverse: Digital Twin of a City in the Alps}},
	author       = {Gallist, Nils and Hagler, Juergen},
	year         = 2023,
	booktitle    = {Proc. of the 22nd International Conference on Mobile and Ubiquitous Multimedia (MUM)},
	location     = {Vienna, Austria,},
	publisher    = {Association for Computing Machinery},
	address      = {New York, NY, USA},
	pages        = {568–570},
	doi          = {10.1145/3626705.3631880},
	isbn         = 9798400709210
}

@incollection{Galvao2024,
	title        = {{Towards a Consensual Definition for Smart Tourism and Smart Tourism Tools}},
	author       = {António Galvão and Fernando Brito e Abreu and João Joanaz de Melo},
	year         = 2024,
	booktitle    = {Smart Life and Smart Life Engineering: Current State and Future Vision},
	publisher    = {Springer Nature},
	series       = {LNBIP},
	pages        = {153--182},
	doi          = {10.1007/978-3-031-75887-4_8},
	issn         = {1865-1356},
	editor       = {Elena Kornyshova and Rébecca Deneckère and Eric Gressier-Soudan and John Murray and Sjaak Brinkkemper},
	chapter      = 8
}

@article{Guzzetti2022,
	title        = {{BIM and Castello Sforzesco in Milan. A particular approach to digitization of the architectural and information heritage}},
	author       = {Guzzetti, F. and Anyabolu, K. L. N. and Biolo, F. and Dell'Orto, R.},
	year         = 2022,
	journal      = {The International Archives of the Photogrammetry, Remote Sensing and Spatial Information Sciences},
	volume       = {XLVI-5/W1-2022},
	pages        = {115--122},
	doi          = {10.5194/isprs-archives-XLVI-5-W1-2022-115-2022}
}

@article{Hazeleger2024,
	title        = {{Digital twins of the Earth with and for humans}},
	author       = {Hazeleger, W. and Aerts, J. P. M. and Bauer, P. and Bierkens, M. F. P. and Camps-Valls, G. and Dekker, M. M. and Doblas-Reyes, F. J. and Eyring, V. and Finkenauer, C. and Grundner, A. and Hachinger, S. and Hall, D. M. and Hartmann, T. and Iglesias-Suarez, F. and Janssens, M. and Jones, E. R. and Kölling, T. and Lees, M. and Lhermitte, S. and van Nieuwpoort, R. V. and Pahker, A.-K. and Pellicer-Valero, O. J. and Pijpers, F. P. and Siibak, A. and Spitzer, J. and Stevens, B. and Vasconcelos, V. V. and Vossepoel, F. C.},
	year         = 2024,
	journal      = {Communications Earth \& Environment},
	volume       = 5,
	number       = 1,
	pages        = 463,
	doi          = {10.1038/s43247-024-01626-x}
}

@article{Huang2024,
	title        = {{Inheritance and Innovation: A Study on the Path of Ancient Village Cultural Tourism Planning and Design}},
	author       = {Huang, Yunyi},
	year         = 2024,
	journal      = {International Journal of Education and Humanities},
	volume       = 17,
	number       = 2,
	pages        = {77--81},
	doi          = {10.54097/22whvz27},
	note         = {https://drpress.org/ojs/index.php/ijeh/article/download/27546/27077}
}

@inproceedings{Kourtit2025,
	title        = {{Methodology and Application of 3D Visualization in Sustainable Cultural Tourism Planning}},
	author       = {Kourtit, Karima and Nijkamp, Peter and Scholten, Henk and van Iersel, Yneke},
	year         = 2025,
	booktitle    = {Proc. of the International Conference on Cultural Tourism Advances},
	publisher    = {Springer Nature Switzerland},
	address      = {Cham},
	pages        = {173--186},
	doi          = {10.1007/978-3-031-65537-1_11},
	isbn         = {978-3-031-65537-1},
	editor       = {Neuts, Bart and Martins, Jo{\~a}o and Ioannides, Marinos}
}

@article{Kumar2022,
	title        = {{Rediscovering the Traditional UNESCO World Heritage Hawamahal through 3D Animation and Immersive Technology}},
	author       = {Kumar, Abhishek and Kumar, Ankit and Raja, Linesh and Singh, Kamred Udham},
	year         = 2023,
	month        = {feb},
	journal      = {Journal on Computing and Cultural Heritage},
	publisher    = {Association for Computing Machinery},
	address      = {New York, NY, USA},
	volume       = 15,
	number       = 4,
	doi          = {10.1145/3524023},
	issn         = {1556-4673},
	issue_date   = {December 2022},
	articleno    = 61,
	numpages     = 34
}

@inproceedings{Li2024,
	title        = {{Design of Smart Tourism Visual Analysis Platform Based on Digital Twin}},
	author       = {Li, Suiqun and Feng, Yidan and Chu, Qikai and Chen, Guangjian},
	year         = 2024,
	booktitle    = {Proc. of the 3rd International Conference on Electrical Engineering, Big Data and Algorithms (EEBDA)},
	pages        = {1300--1303},
	doi          = {10.1109/EEBDA60612.2024.10485670},
	organization = {IEEE}
}

@article{Litavniece2023,
	title        = {{Digital twin: an approach to enhancing tourism competitiveness}},
	author       = {Litavniece, Lienite and Kodors, Sergejs and Adamoniene, R{\={u}}ta and Kijasko, Jelena},
	year         = 2023,
	month        = {Jan},
	day          = {01},
	journal      = {Worldwide Hospitality and Tourism Themes},
	publisher    = {Emerald Publishing Limited},
	volume       = 15,
	number       = 5,
	pages        = {538--548},
	doi          = {10.1108/WHATT-06-2023-0074},
	issn         = {1755-4217}
}

@inproceedings{Logothetis2023,
	title        = {{Towards a Digital Twin Implementation of Eastern Crete: An Educational Approach}},
	author       = {Logothetis, Ilias and Mari, Ioanna and Vidakis, Nikolas},
	year         = 2023,
	booktitle    = {Proc. of the International Conference on Extended Reality (XR)},
	location     = {Lecce, Italy},
	publisher    = {Springer-Verlag},
	address      = {Berlin, Heidelberg},
	pages        = {255–268},
	doi          = {10.1007/978-3-031-43401-3_17},
	isbn         = {978-3-031-43400-6},
	numpages     = 14
}

@article{LopezGonzalez2023,
	title        = {{The Integration of HBIM-SIG in the Development of a Virtual Itinerary in a Historical Centre}},
	author       = {López-González, Concepción and García-Valldecabres, Jorge},
	year         = 2023,
	journal      = {Sustainability},
	volume       = 15,
	number       = 18,
	doi          = {10.3390/su151813931},
	issn         = {2071-1050},
	article-number = 13931
}

@inbook{Menaguale2023,
	title        = {{Digital Twin and Cultural Heritage -- The Future of Society Built on History and Art}},
	author       = {Menaguale, Olivia},
	year         = 2023,
	booktitle    = {The Digital Twin},
	publisher    = {Springer International Publishing},
	address      = {Cham},
	pages        = {1081--1111},
	doi          = {10.1007/978-3-031-21343-4_34},
	isbn         = {978-3-031-21343-4},
	editor       = {Crespi, Noel and Drobot, Adam T. and Minerva, Roberto}
}

@book{NASA2012,
	title        = {{NASA space technology roadmaps and priorities: restoring NASA's technological edge and paving the way for a new era in space}},
	author       = {National Research Council and Division on Engineering and Physical Sciences and Aeronautics and Space Engineering Board and Steering Committee for NASA Technology Roadmaps},
	year         = 2012,
	publisher    = {National Academies Press}
}

@inproceedings{Nguyen2022,
	title        = {{Integration of H-BIM, Virtual Reality, and Augmented Reality in Digital Twin Era - A Case Study in Cultural Heritage}},
	author       = {Nguyen, Thu Anh and Do, Sy Tien and Pham, Truong-An and Nguyen, Dai Huu and Tamura, Hiroshi},
	year         = 2022,
	booktitle    = {Proc. of the Second International Conference on Sustainable Civil Engineering and Architecturen (ICSCEA)},
	publisher    = {Springer Nature Singapore},
	address      = {Singapore},
	pages        = {303--312},
	doi          = {10.1007/978-981-19-3303-5_24},
	isbn         = {978-981-19-3303-5},
	editor       = {Reddy, J. N. and Wang, Chien Ming and Luong, Van Hai and Le, Anh Tuan}
}

@article{Planu2024,
	title        = {{Digital territorial assets: vocational drivers’ representation for Finiq Municipality's challenge of change}},
	author       = {Planu, Fabio},
	year         = 2024,
	journal      = {The Scientific Journal of the Observatory of Mediterranean Basin},
	volume       = {},
	number       = 9,
	pages        = {174--185},
	doi          = {10.37199/o41009114}
}

@incollection{Prabawati2024,
	title        = {{Digital Twin and Tourism: Recreating and Reimagining Tourist Experience by Interconnecting Physical and Virtual Systems}},
	author       = {Prabawati, Andhika Galuh and Tamtama, Gabriel Indra Widi and Santoso, Halim Budi},
	year         = 2024,
	booktitle    = {Tourism and Hospitality for Sustainable Development: Volume Three: Implications for Customers and Employees of Tourism Businesses},
	publisher    = {Springer Nature Switzerland},
	address      = {Cham},
	pages        = {45--65},
	doi          = {10.1007/978-3-031-63077-4_3},
	isbn         = {978-3-031-63077-4},
	editor       = {Ndhlovu, Emmanuel and Dube, Kaitano and Makuyana, Tawanda}
}

@article{Rahmadian2023,
	title        = {{Digital twins, big data governance, and sustainable tourism}},
	author       = {Rahmadian, Eko and Feitosa, Daniel and Virantina, Yulia},
	year         = 2023,
	month        = {Nov},
	day          = 16,
	journal      = {Ethics and Information Technology},
	volume       = 25,
	number       = 4,
	pages        = 61,
	doi          = {10.1007/s10676-023-09730-w},
	issn         = {1572-8439}
}

@incollection{Rahmadian2023Governing,
	title        = {{Governing Digital Twin technology for smart and sustainable tourism: a case study in applying a documentation framework for architecture decisions}},
	author       = {Eko Rahmadian and Daniel Feitosa and Andrej Zwitter},
	year         = 2023,
	booktitle    = {{Handbook on the Politics and Governance of Big Data and Artificial Intelligence}},
	publisher    = {Edward Elgar Publishing},
	pages        = {105--137},
	doi          = {10.4337/9781800887374.00014},
	editor       = {Andrej Zwitter and Oskar J. Gstrein},
	chapter      = 4
}

@article{Rolph2024,
	title        = {{Prototype Digital Twin: Recreation and biodiversity cultural ecosystem services}},
	author       = {Rolph, Simon and Andrews, Christopher and Carbone, Dylan and Lopez Gordillo, Julian and Martinovi{\v{c}}, Tom{\'a}{\v{s}} and Oostervink, Nick and Pleiter, Dirk and Watkins, John and Wohner, Christoph and Bolton, Will and others},
	year         = 2024,
	journal      = {Research Ideas and Outcomes},
	publisher    = {Pensoft},
	volume       = 10,
	doi          = {10.3897/rio.10.e125450}
}

@inproceedings{Sang2022,
	title        = {{Digital Twin Platform Design for Zhejiang Rural Cultural Tourism Based on Unreal Engine}},
	author       = {Sang, Feiyang and Wu, Hui and Liu, Zhe and Fang, Shenguo},
	year         = 2022,
	booktitle    = {Proc. of the International Conference on Culture-Oriented Science and Technology (CoST)},
	pages        = {274--278},
	doi          = {10.1109/CoST57098.2022.00063}
}

@incollection{Seungyoub2021,
	title        = {{Science Tour and Business Model Using Digital Twin-Based Augmented Reality}},
	author       = {Ssin, Seungyoub and Suh, Minjeong and Lee, Jongwook and Jung, Timothy and Woo, Woontack},
	year         = 2021,
	booktitle    = {Augmented Reality and Virtual Reality},
	publisher    = {Springer International Publishing},
	address      = {Cham},
	pages        = {267--276},
	doi          = {10.1007/978-3-030-68086-2_20},
	isbn         = {978-3-030-68086-2},
	editor       = {tom Dieck, M. Claudia and Jung, Timothy H. and Loureiro, Sandra M. C.}
}

@article{Tang2024,
	title        = {{Digitalisation of asset management in Chinese temples in Hong Kong}},
	author       = {Tang, Whai Tak and Siu, Terry and Villupuram, Shrisankaraan S and Lee, Thomas and Ji, Maggie},
	year         = 2024,
	journal      = {HKIE Transactions},
	volume       = 30,
	number       = 3,
	pages        = {34--43},
	doi          = {10.33430/v30n3thie-2022-0053}
}

@book{Wan2023,
	title        = {{Digital twins for smart cities: Conceptualisation, challenges and practices}},
	author       = {Wan, Li and Nochta, Timea and Tang, Junqing and Schooling, Jennifer},
	year         = 2023,
	publisher    = {ICE Publishing}
}

@book{Willcox2024,
	title        = {{Foundational Research Gaps and Future Directions for Digital Twins}},
	author       = {Karen E. Willcox and Derek Bingham and Caroline Chung and Julianne Chung and Carolina Cruz-Neira and Conrad J. Grant and James L. Kinter and Ruby Leung and Parviz Moin and Lucila Ohno-Machado and Colin J. Parris and Irene Qualters and Ines Thiele and Conrad Tucker and Rebecca Willett and Xinyue Ye},
	year         = 2024,
	publisher    = {The National Academies Press},
	address      = {Washington, DC},
	doi          = {10.17226/26894},
	isbn         = {978-0-309-70042-9}
}

@article{Yang2021,
	title        = {{Research on Guilin Smart Cultural Tourism Service Scenario based on Digital Twin Technology}},
	author       = {Yang, Jingran and Peng, Yuyuan and Liang, Hongjin},
	year         = 2021,
	journal      = {BCP Business \& Management},
	volume       = 13,
	pages        = {19--24},
	doi          = {10.54691/bcpbm.v13i.51}
}

@article{Zhang2022,
	title        = {{Metaverse for Cultural Heritages}},
	author       = {Zhang, Xiao and Yang, Deling and Yow, Cheun Hoe and Huang, Lihui and Wu, Xiaoqun and Huang, Xijun and Guo, Jia and Zhou, Shujun and Cai, Yiyu},
	year         = 2022,
	journal      = {Electronics},
	volume       = 11,
	number       = 22,
	doi          = {10.3390/electronics11223730},
	issn         = {2079-9292},
	article-number = 3730
}

@article{Zhang2023,
	title        = {{Digital Twin System for Tourist Attractions Based on 3D GIS Technology}},
	author       = {Zhang, Yuxing},
	year         = 2023,
	journal      = {Advances in Computer and Communication},
	volume       = 4,
	number       = 5,
	pages        = {324--327},
	doi          = {10.26855/acc.2023.10.011}
}

@article{Zhao2022,
	title        = {{Application of Digital Twin Combined with Artificial Intelligence and 5G Technology in the Art Design of Digital Museums}},
	author       = {Zhao, Jing and Guo, Lei and Li, Yueqiao},
	year         = 2022,
	month        = {JUN 2},
	journal      = {Wireless Communications and Mobile Computing},
	publisher    = {Wiley-Hindawi},
	address      = {London, England},
	volume       = 2022,
	doi          = {10.1155/2022/8214514},
	issn         = {1530-8669}
}

@inproceedings{Zhou2023,
	title        = {{Research and Practice of the New Mode of Tourism Service by Digital Twin Technology Based on UAV-Taking Sanjiang Gaoyou Village As an Example}},
	author       = {Wei Zhou and Zhifeng Mao},
	year         = 2023,
	booktitle    = {Proc. of the 2nd International Conference on Intelligent Design and Innovative Technology (ICIDIT)},
	publisher    = {Atlantis Press},
	pages        = {471--477},
	doi          = {10.2991/978-94-6463-266-8_51},
	isbn         = {978-94-6463-266-8},
	issn         = {2589-4919}
}



\end{document}